\documentclass[twocolumn]{aastex61}

\newcommand{\rhoFRB}{\rho_\mathrm{FRB}}
\newcommand{\DMex}{{\rm DM_{EX}}}
\newcommand{\DMigm}{{\rm DM_{IGM}}}
\newcommand{\DMism}{{\rm DM_{ISM}}}
\newcommand{\DMhalo}{{\rm DM_{halo}}}
\newcommand{\DMhost}{{\rm DM_{host}}}
\newcommand{\DMhostmed}{{\rm DM_{host,med}}}

\newcommand{\Sth}{S_{\nu,{\rm th}}}

\newcommand{\Lo}{L_{\nu, 0}}
\newcommand{\Pks}{P_{\rm KS}} 
\newcommand{\betast}{{\beta_{\rm stat}}}

\received{\today}
\revised{\today}
\accepted{\today}
\submitjournal{ApJ}

\shorttitle{Observational constraints on DM components within FRB host galaxies}
\shortauthors{Niino}

\begin{document}

\title{
Dispersion measure components within host galaxies of Fast Radio Bursts: 
observational constraints from statistical properties of FRBs
}

\correspondingauthor{Yuu Niino}
\email{yuuniino@ioa.s.u-tokyo.ac.jp}

\author[0000-0002-0786-7307]{Yuu Niino}
\affil{Research Center for the Early Universe, Graduate School of Science, 
        University of Tokyo, Bunkyo-ku, Tokyo 113-0033, Japan}
\affil{Institute of Astronomy, Graduate School of Science, 
        University of Tokyo, 2-21-1, Osawa, Mitaka, Tokyo 181-0015, Japan}

\begin{abstract}
Dispersion measure (DM) of Fast Radio Bursts (FRBs) 
are commonly used as a indicator of distance 
assuming that DM in excess of the expected amount 
within the Milky Way in the direction of each FRB  
arise mostly from the inter-galactic medium. 
However, the assumption might not be true if, for example, 
most FRB progenitors are embedded 
in ionized circumstellar material (CSM, e.g. supernova remnant). 
In this study, we jointly analyze distributions of DM, flux density, 
and fluence of the FRB samples observed by the Parkes telescope 
and the Australian Square Kilometre Array Pathfinder (ASKAP) 
using analytical models of FRBs, 
to constrain fractions of various DM components that shape 
the overall DM distribution and emission properties of FRBs. 
Comparing the model predictions with the observations 
we find that the typical amount of DM in each FRB host galaxy 
is $\sim 120$ cm$^{-3}$pc which is naturally explained as a combination 
of interstellar medium (ISM) and halo of an ordinary galaxy, 
without additional contribution from ionized CSM 
that is directly associated with an FRB progenitor. 
Furthermore, we also find that observed flux densities 
of FRBs do not statistically suffer strong $K$-correction, 
i.e. the typical luminosity density of FRBs does not significantly change 
within the range of emitting frequency $\nu_{\rm rest} \sim$ 1--4 GHz. 
\end{abstract}


\section{Introduction} 
\label{sec:intro}

A Fast Radio Burst (FRB) is a transient astronomical 
object observed at $\sim$ 1 GHz frequency with a typical duration 
of several milliseconds, whose origin is not yet known 
\citep[e.g.,][]{Lorimer:2007a, Keane:2012a, Thornton:2013a}. 
More than a hundred FRB sources have been discovered so far. 
Roughly 20 FRB sources are known to produce bursts repeatedly (repeating FRBs), 
while other FRB sources do not show any repetition (non-repeating FRBs), 
implying a possibility that they are different populations 
of astronomical objects \citep{Palaniswamy:2018a}. 

FRBs have large dispersion measures 
(column density of free electrons along a line of sight 
which is measured with delay of pulse arrival time 
as a function of frequency, hereafter DMs) 
that exceed the expected amounts within the Milky Way (MW) in their direction. 
Their large DMs suggest that FRBs are extragalactic objects. 
Although various theoretical models have been proposed 
\citep[e.g.,][see \citeauthor{Platts:2019a}~\citeyear{Platts:2019a} 
for a recent review]{Totani:2013a, Kashiyama:2013a, 
Popov:2013a, Falcke:2014a, Cordes:2016b, Zhang:2017a}, 
observational evidence that confirms or rejects those models is still lacking. 

Most of the currently known FRBs have been discovered by widefield 
radio telescopes with typical localization accuracy of $\gtrsim$ 10 arcmin, 
and hence it is challenging to identify their counterparts  or host galaxies in most 
of the cases \citep[e.g.,][]{Petroff:2015a, DeLaunay:2016a, Niino:2018a, Tominaga:2018a}. 
Currently, identifications of FRB host galaxies, 
and hence distance measurements that are independent of DM, 
have been achieved only for 5 FRBs \citep[2 repeating and 3 non-repeating FRBs 
at redshifts $z \sim 0.03$--0.7,][]{Tendulkar:2017a, 
Bannister:2019a, Ravi:2019a, Prochaska:2019b, Marcote:2020a}. 
Distances of other FRBs are estimated from their DMs assuming 
that the DMs in excess of the expected MW component ($\DMex$) 
arise mostly from the inter-galactic medium (IGM), 
and considered to be widely distributed over a redshift range $z \sim 0.1$--2.5. 
However the actual distances of FRBs can be shorter than the estimations 
if significant fraction of the DMs arise from other ionized gas components than the IGM. 

As locations of FRBs are not known in most of the cases, 
statistical distributions of observed $\DMex$ and flux density (or fluence) 
are important clues to understand the nature of FRBs 
\citep[e.g.,][]{Dolag:2015a, Katz:2016a, Caleb:2016a}. 
Analyzing the statistical properties 
of FRBs discovered by the Parkes radio telescope, 
\citet[hereafter N18]{Niino:2018b} showed that the observed properties 
are better explained if FRBs are at cosmological distances 
and the cosmic FRB rate density [$\rhoFRB (z)$] 
increases with redshift resembling the cosmic star formation history (CSFH), 
while a model in which FRBs originate in the local universe 
(i.e., $\DMex$ is dominated by non-IGM component) is disfavored. 
However, quantitative constraint on the fraction 
of the IGM and non-IGM components in $\DMex$ of FRBs 
was not obtained by the analysis in N18. 

Recently, an energetic radio burst from a Galactic magnetar, 
SGR~1935+2154, was observed \citep{CHIME:2020a, Bochenek:2020a}. 
The magnetar radio burst emitted $\sim 10^{35}$ erg during its $\sim 1$ ms duration. 
This luminosity is comparable to $\sim$ 1/30 
of the faintest extragalactic FRB ever observed \citep[a burst from a repeating 
FRB~180916.J0158+65][]{Marcote:2020a}, or $\sim 10^{-4}$ of typical FRBs. 
Whether all (or majority of) FRBs are similar phenomena 
to the burst from SGR~1935+2154 is still a matter of debate. 
It is possible to reconcile the inferred event rate of magnetar radio bursts 
like that from SGR~1935+2154 with the luminosity function (LF) of 
extragalactic FRBs \citep{Margalit:2020a, Lu:2020a}. 
However, \citet{Margalit:2020a} also pointed out 
that the typical distance of extragalactic FRBs would be 
shorter than that estimated from $\DMex$, 
if the FRB LF is directly connected to the inferred 
event rate of magnetar radio bursts in its faint-end. 

Previous investigations of FRB LF have been conducted 
assuming that contributions of non-IGM components 
to $\DMex$ do not significantly exceed the amount expected 
from diffuse interstellar medium (ISM) of an FRB host galaxy 
which is usually smaller than the IGM component 
\citep[N18; ][]{Luo:2018a, Luo:2020a, Lu:2019a}. 
However, the assumption might not be true if most FRB progenitors 
are embedded in ionized circumstellar material 
\citep[CSM, e.g. supernova remnant, pulsar wind nebula, HII region, 
see][]{Kokubo:2017a, Piro:2017a}. 
It is essential to unveil actual distances and luminosities of FRBs, 
to understand the nature of FRBs and clarify their relation to magnetar radio bursts. 
In this study, we jointly analyze the statistical properties 
of the FRB samples observed by the Parkes telescope 
and the Australian Square Kilometre Array Pathfinder (ASKAP), 
to put constraints on non-IGM DM components associated 
with FRB progenitors based on the observational data. 

In Section~\ref{sec:sample}, we describe the datasets 
that we use to constrain the properties of FRBs. 
In Section~\ref{sec:models}, we describe our model of FRB population. 
In Section~\ref{sec:results}, we discuss constraints 
on the typical luminosity of FRBs and the amount 
of non-IGM DM components from the $\DMex$ distributions
of the observed FRB samples. 
In Section~\ref{sec:kcorrection}, we discuss how the constraints 
are affected by spectra of FRBs statistically (i.e., effects of $K$-correction). 
In Section~\ref{sec:flux}, we discuss constraints obtained from 
the distribution functions of flux densities and fluences of the FRB samples.  
We summarize our conclusions in section~\ref{sec:conc}. 
Throughout this paper, we assume the fiducial cosmology 
with $\Omega_{\Lambda}=0.7$, $\Omega_{m}=0.3$, 
and $H_0=$ 70 km s$^{-1}$ Mpc$^{-1}$. 

\section{FRB datasets} 
\label{sec:sample} 

To constrain the properties of FRB, we use the sample of FRBs 
discovered by the Parkes telescope and ASKAP. 
The properties of the observed FRBs are collected 
from the FRBCAT\footnote{http://frbcat.org} database \citep{Petroff:2016a}, 
as reported at the beginning of September 2019. 
We exclude FRB180923, the last Parkes FRB before the date of data collection, 
from the Parkes sample because its peak flux density 
which we use in our analysis is not reported. 
We also exclude FRBs discovered by ASKAP 
in 2019 (FRB~190711, and 190714) 
from the ASKAP sample for the same reason. 
We note that there are four outlier FRBs in the Parkes sample 
compared to the overall $S_\nu$--DM distribution of the sample, 
which have extremely bright flux density and small DM, 
and we exclude these FRBs (FRB~010724, 110214, 150807, 180309) from our analysis. 
In total, the Parkes sample includes 24 FRBs between FRB~010125 and 180714, 
and the ASKAP sample includes 26 FRBs between FRB~170107 and 180924. 

\section{Models} 
\label{sec:models} 

\subsection{$\rhoFRB$ and LF} 

\begin{figure}
\plotone{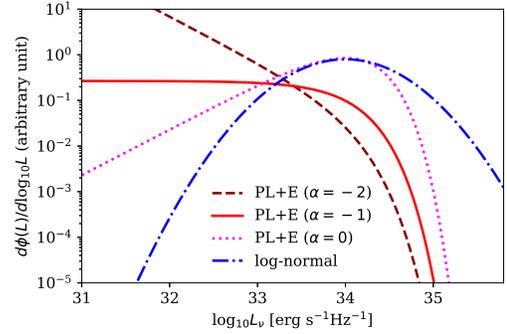}
\caption{
The LF models considered in this study. 
The dashed, solid, and dotted lines represent the power-law plus exponential cutoff 
(PL+E) models with $\alpha = -2, -1$, and 0, respectively. 
The dot-dashed line represents the log-normal model. 
$\Lo = 10^{34}$ erg s$^{-1}$Hz$^{-1}$ in this figure. 
\label{fig:LF}
}
\end{figure}

Investigating the distributions of DM and apparent flux density 
of FRBs discovered by the Parkes radio telescope, 
N18 showed that $\rhoFRB$ increases with redshift resembling CSFH, 
and an LF model with a bright-end cutoff 
at log$_{10}L_\nu$ [erg s$^{-1}$Hz$^{-1}$] $\sim$ 34 
are favored to reproduce observations, 
while faint-end of the LF is not well constrained. 
The existence of the bright-end cutoff in FRB LF is 
also independently shown by \citet{Luo:2018a}. 
In this study, we assume that $\rhoFRB (z)$ is proportional 
to CSFH as derived by \citet{Madau:2014a}, 
and use the following LF models to examine how our results 
are affected by the difference of the faint end of the FRB LF (Figure~\ref{fig:LF}); 
\begin{itemize} 
\item Power-law distribution function with index $\alpha$,  
and exponential cutoff in the bright-end above $\Lo$ (PL+E):  
\begin{equation} 
\frac{d\phi}{dL_\nu} \propto L_\nu^\alpha exp(-\frac{L_\nu}{\Lo}). 
\end{equation} 
We consider the cases with  $\alpha = -2, -1$, and 0 in this study 
($\alpha = -1$ as a baseline and $\alpha = -2, 0$ as steep and flat variants).  
\item Log-normal distribution with median $\Lo$: 
\begin{equation} 
\frac{d\phi}{dL_\nu} \propto \frac{{\rm log_{10}}e}{\sqrt{2\pi} \sigma L_\nu}
{\rm exp}(-\frac{({\rm log_{10}}L_\nu-{\rm log_{10}}\Lo)^2}{2\sigma^2}). 
\end{equation} 
We assume $\sigma = 0.5$ dex in this study. 
\end{itemize} 
$\Lo$ in each LF model is a free parameter which will be constrained 
by comparing model predictions with observations. 

\subsection{Receiver efficiency, propagation effect, and $K$-correction} 

It should be noted that the efficiency 
of the Parkes multi-beam receiver \citep{Staveley-Smith:1996a} 
largely varies within its beam, 
and the reported flux densities and fluences are converted 
from the observed signal assuming the receiver efficiency 
at the beam center, and thus effectively are lower-limits.  

To account for the variation of the receiver efficiency, 
N18 computed the probability distribution function (PDF) 
of receiver efficiency assuming the beam shape 
of the Parkes multi-beam receiver is represented by an Airy disc, 
and convoluted the flux density PDF of the FRB model with the receiver efficiency PDF. 
We follow the method of N18 when we compare 
our model predictions with the Parkes sample of observed FRBs. 
On the other hand, the ASKAP phased array feed receivers sample 
the focal plane almost uniformly \citep{Bannister:2017a, Shannon:2018a}, 
and hence we do not use the receiver efficiency PDF model 
when we discuss the observed properties of the ASKAP sample. 

Observed flux density of an FRB is also affected 
by various propagation effects between the source and the observer. 
Scattering of an FRB signal suppress FRB flux density by pulse broadening, 
on the other hand, scintillation and plasma lensing may also enhance FRB flux density 
\citep[e.g.,][]{Hassall:2013a, Cordes:2016a, Cordes:2017a}. 
Following N18, we treat PDF of propagation effects 
as included in the FRB LF rather than trying to separate 
intrinsic luminosity of an FRB from propagation effects. 

$K$-correction is also an important effect when we consider 
observed flux densities of objects at cosmological distances. 
In this study, we consider the cace in which the $K$-correction factor 
$\kappa_\nu(z) = L_\nu(\nu_{\rm rest})/L_\nu(\nu_{\rm obs}) = 1$ 
as a baseline model in Section~\ref{sec:results}, 
and discuss how our the results are affected 
by $K$-correction in Section~\ref{sec:kcorrection}. 
Here $\kappa_\nu(z) = 1$ means that the typical spectral index 
of FRBs is 0  in a statistical meaning ($\betast = 0$), 
or that the FRB LF is not changed with rest frame frequency in the range 
of $\nu_{\rm rest} \sim$ 1--4 GHz which is covered by the current sample, 
but not that most FRBs have a flat spectrum. 

\subsection{The detection threshold}

To make model predictions of observed FRB properties, 
we need to determine a detection threshold for model FRBs. 
In N18, we considered a model FRB as detected 
when its flux density exceeds a threshold value, $S_\nu \geq \Sth$. 
Detectability of an FRB is affected not only by its flux density 
but also on the pulse width in reality \citep[and hence the fluence ($F_\nu$),][]{Keane:2015a}. 
However, N18 pointed out that signal-to-noise ratio (S/N) of FRB detections 
in the Parkes sample empirically correlates well with flux density, 
and the faint-end of the flux density distribution of the Parkes sample is sharply cut. 
These facts suggest that $S_\nu$ is effectively a good proxy for S/N. 
In this study, we assume the threshold flux density 
of $\Sth = 0.4$ Jy and 15 Jy for the Parkes and ASKAP samples, 
based on the faint end of the $S_\nu$ distributions 
of the two samples which we show in Section~\ref{sec:flux}. 

\subsection{DM computation}

We consider DM of an FRB as a summation of 4 components, 
DM = $\DMism + \DMhalo + \DMigm + \DMhost$, 
where $\DMism$ and $\DMhalo$ are the DM components 
associated with the ISM and the halo of MW, 
$\DMigm$ arises from the IGM between the host galaxy of the FRB and MW, 
and $\DMhost$ arises from ionized gas within the host galaxy. 
We note that $\DMhost$ includes DM components that arise 
from the galaxy scale ISM, the host galaxy halo, and possible CSM 
that is directly associated with the progenitor of the FRB 
(e.g. supernova remnant, pulsar wind nebula, HII region). 

In the FRBCAT database, excess of observed DM beyond $\DMism$ 
($\DMex = \DMhalo + \DMigm + \DMhost$) is reported for each event 
assuming the NE2001 model of free electrons in the Galactic ISM \citep{Cordes:2002a}, 
and we compare these values to the model predictions. 
Some previous studies have independently shown that $\DMhalo$ is typically 
$\sim 50$ cm$^{-3}$pc \citep{Dolag:2015a, Prochaska:2019a, Yamasaki:2020a}. 
In this study, we assume $\DMhalo = 30$ cm$^{-3}$pc for any FRB following \citet{Dolag:2015a}. 

$\DMigm$ is determined by the distance 
between the FRB host galaxy and MW, 
i.e., redshift of the FRB \citep[e.g.,][]{Ioka:2003a, Inoue:2004a}. 
We use the same formalism of $\DMigm$ as in N18. 
In the redshift range discussed in this paper, the formalism 
can be naively approximated as $\DMigm \sim 1000z$ cm$^{-3}$pc. 

We assume $\DMhost$ follows a log-normal distribution with $\sigma = 0.2$ dex, 
motivated by theoretical models of $\DMhost$ distribution 
in a disk galaxy \citep{Xu:2015a, Walker:2018a, Luo:2018a}. 
The median value of the distribution, $\DMhostmed$, 
is a free parameter we will constrain in the following sections. 
We note that the $\DMex$ distributions of the Parkes and ASKAP samples 
can be also approximated by log-normal distributions with $\sigma = 0.2$ dex, 
and hence if $\DMhost$ occupies a large fraction of observed $\DMex$, 
$\DMhost$ must follow a PDF that resembles 
the log-normal distribution with $\sigma = 0.2$ dex to explain observations. 

\section{Constraining the characteristic luminosity of FRBs 
        and the amount of DM within their host galaxies} 
\label{sec:results} 

\subsection{Fitting the DM distributions without DM components associated with FRB sources} 
\label{sec:1Dfitting} 

The FRB models described in Section~\ref{sec:models} 
have two free parameters, $\Lo$ and $\DMhostmed$, 
and we will constrain these parameters by comparing predicted $\DMex$ distributions 
with the $\DMex$ distributions of the observed samples. 
First, we perform $\DMex$ distribution fitting using one parameter, $\Lo$, 
assuming that contribution of an FRB host galaxy to $\DMex$ is negligible. 

\begin{figure*}
\plotone{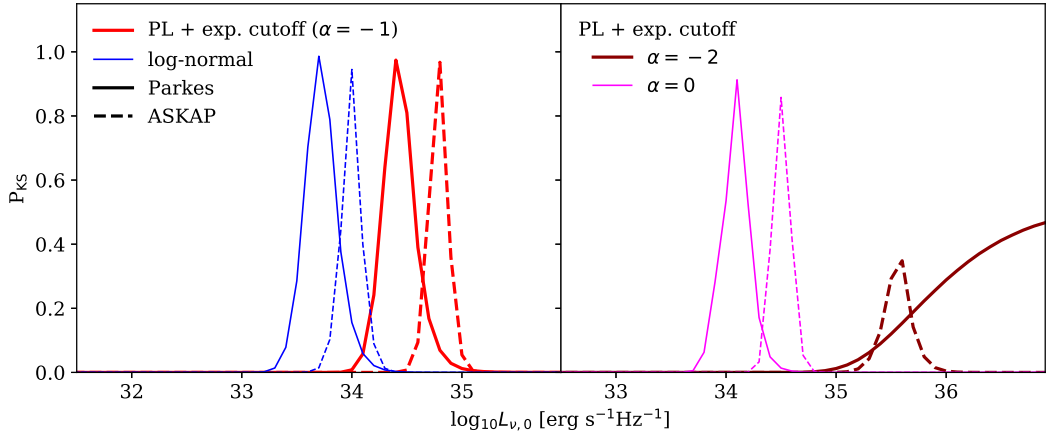}
\caption{
$\Pks$ likelihood between the model 
and the observed $\DMex$ distributions as a function of $\Lo$. 
The solid and dashed lines indicate the results 
for the Parkes and ASKAP samples, respectively. 
Lines with different thickness (different color in the colored version) 
represent results with different LM models. 
\label{fig:DMfit_noHost}
}
\end{figure*}

\begin{figure*}
\plottwo{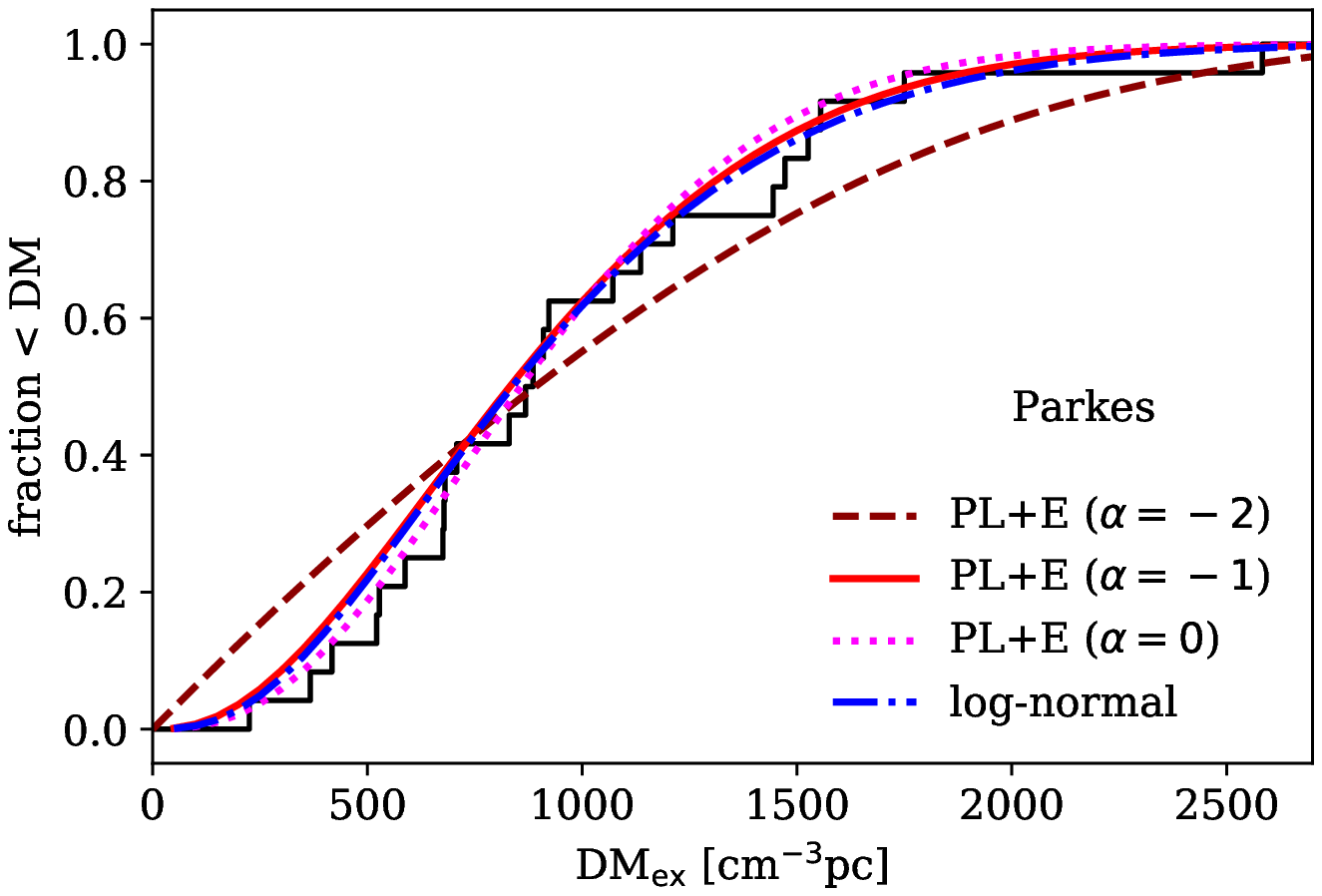}{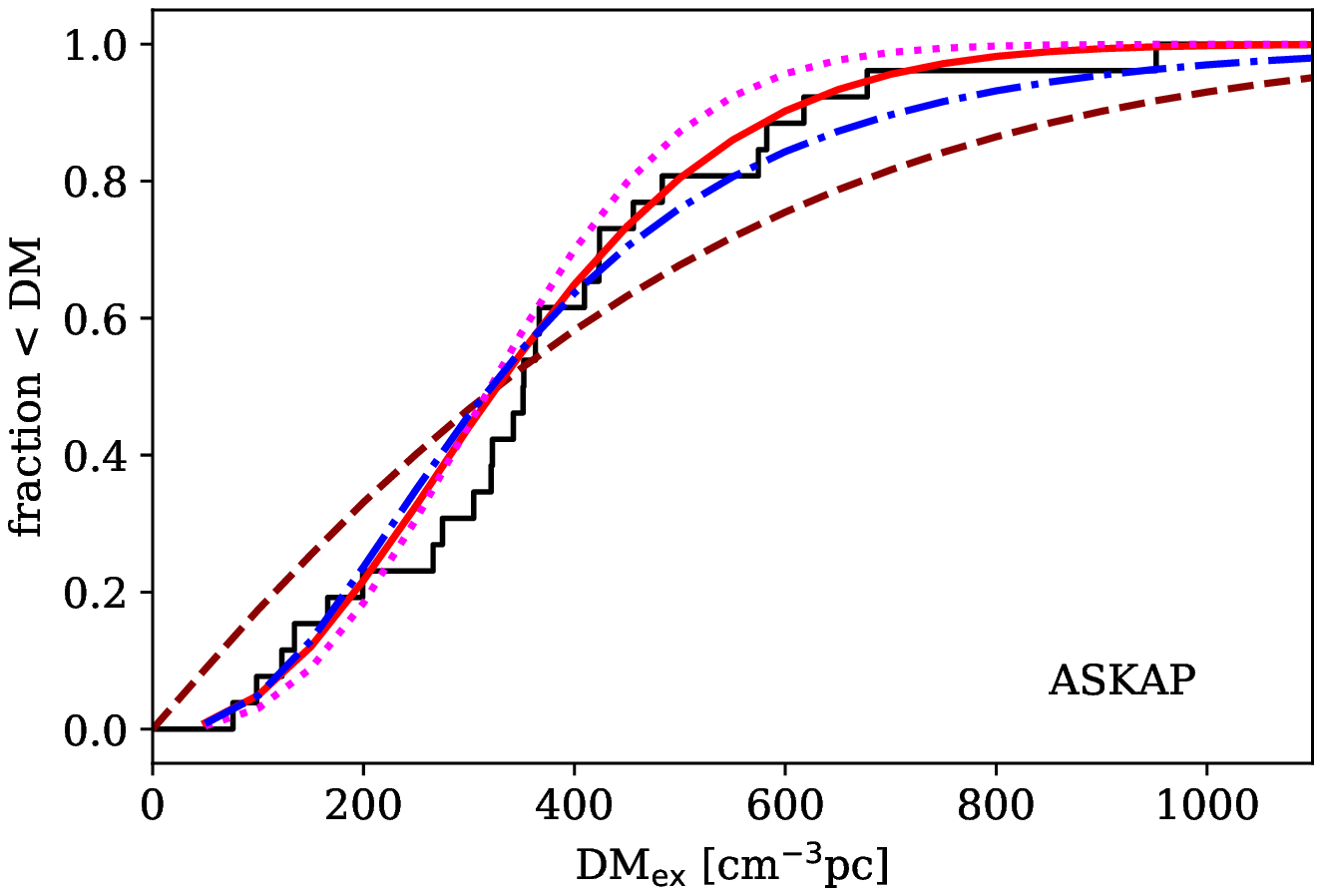} 
\caption{
The cumulative $\DMex$ distributions of the observed FRB samples (histogram), 
and the best fit model distributions to the observations with $\DMhost=0$. 
The left and right panels shows the distributions 
of the Parkes and ASKAP samples, respectively. 
The dashed, solid, and dotted lines represent the model distributions 
with the PL+E LF models with $\alpha = -1.0, -2.0$, and 0.0, respectively. 
The dot-dashed line represents the distribution with the log-normal LF model.  
\label{fig:bestfitmodels}
}
\end{figure*}

The goodness of fit is evaluated by the Kolmogorov-Smirnov (KS) test. 
Figure~\ref{fig:DMfit_noHost} shows the KS test probability ($\Pks$) 
that the observed sample can arise from the model distribution as a function of $\Lo$, 
and Figure~\ref{fig:bestfitmodels} shows the best-fit $\DMex$ distributions. 
With any of the LF models except the steep PL+E ($\alpha = -2$), 
the best-fit $\Lo$ is larger for the ASKAP sample. 
On the other hand, $\Lo$ is poorly constrained for the Parkes sample 
in the case of the steep PL+E model of LF. 

\subsection{Fitting the DM distributions with DM components associated with FRB sources} 
\label{sec:2Dfitting} 

The observing frequencies of the Parkes telescope 
and ASKAP are similar to each other, 
and hence we consider that the best-fitting parameters 
that result form the fittings to the two samples should be the same. 
Here we perform the $\DMex$ distribution fitting 
using two parameters ($\Lo$ and $\DMhostmed$) 
to find parameter sets that reproduce the $\DMex$ distributions 
of the two samples simultaneously. 

\begin{figure*} 
\plotone{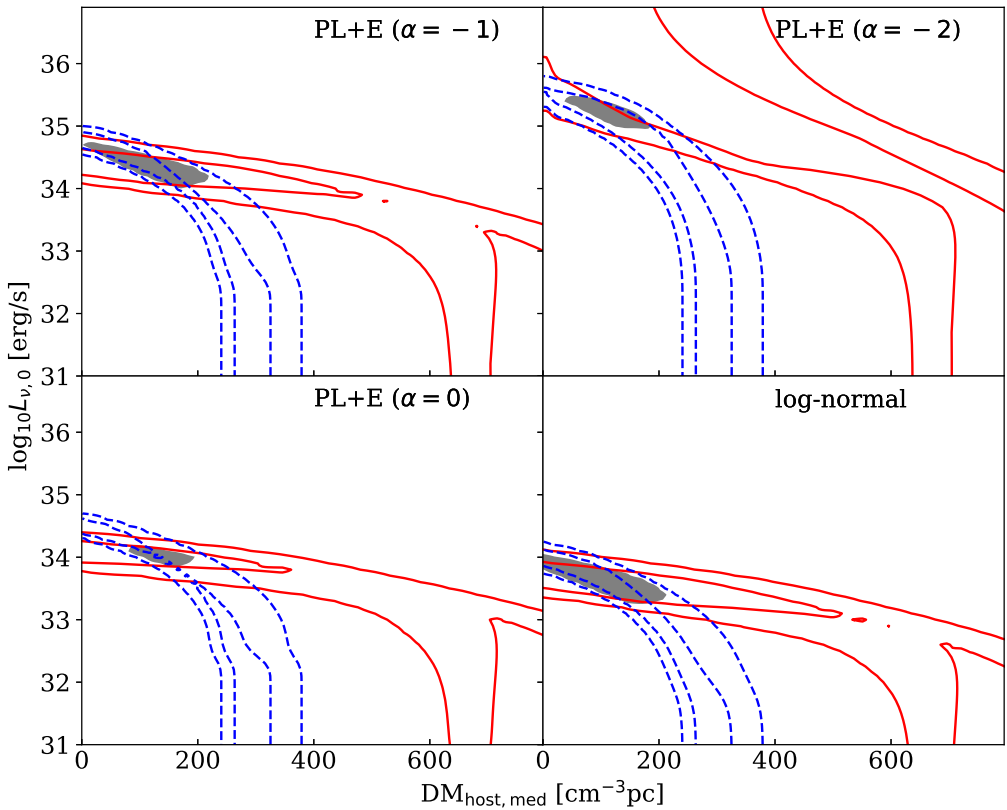}
\caption{
$\Pks$ likelihood map on the parameter plane of $\DMhostmed$ and $\Lo$. 
The solid and dashed contours represent $\Pks$ (at 0.05 and 0.32) 
for the Parkes and ASKAP samples, respectively. 
Different panels show the $\Pks$ likelihood 
for the four different LF models. 
The gray shaded regions indicates 
$P_{\rm KS,joint} = P_{\rm KS,Parkes} \times P_{\rm KS,ASKAP} > 0.1$. 
\label{fig:2Dfit_noK}
}
\end{figure*}

The acceptable ranges of the parameters are shown 
as contours of $\Pks$ for each of the observed samples in Figure~\ref{fig:2Dfit_noK}. 
The parameter ranges that can reproduce the $\DMex$ distributions 
of the Parkes and ASKAP samples at a same time are also indicated
($P_{\rm KS,joint} = P_{\rm KS,Parkes} \times P_{\rm KS,ASKAP} > 0.1$). 
The $\DMex$ distributions of the two samples 
can be consistently explained when $\DMhostmed \sim 120$ cm$^{-3}$pc, 
while $\DMhostmed \gtrsim 200$ cm$^{-3}$pc is disfavored with any of the LF models

It has been shown that contribution from ISM of a MW like host galaxy to observed DM 
of an FRB is $\sim 100$ cm$^{-3}$pc \citep{Xu:2015a, Walker:2018a, Luo:2018a}. 
The predicted typical contribution of a FRB host galaxy to observed DM, 
$\DMhostmed \sim 120$ cm$^{-3}$pc, can be 
naturally explained as that arise from the ISM and the halo of the host galaxy, 
and hence suggests that a DM component that arise 
from CSM that is directly associated with an FRB progenitor 
is typically small ($< 80$ cm$^{-3}$pc). 

\section{Effects of K-correction} 
\label{sec:kcorrection} 

\begin{figure*} 
\plotone{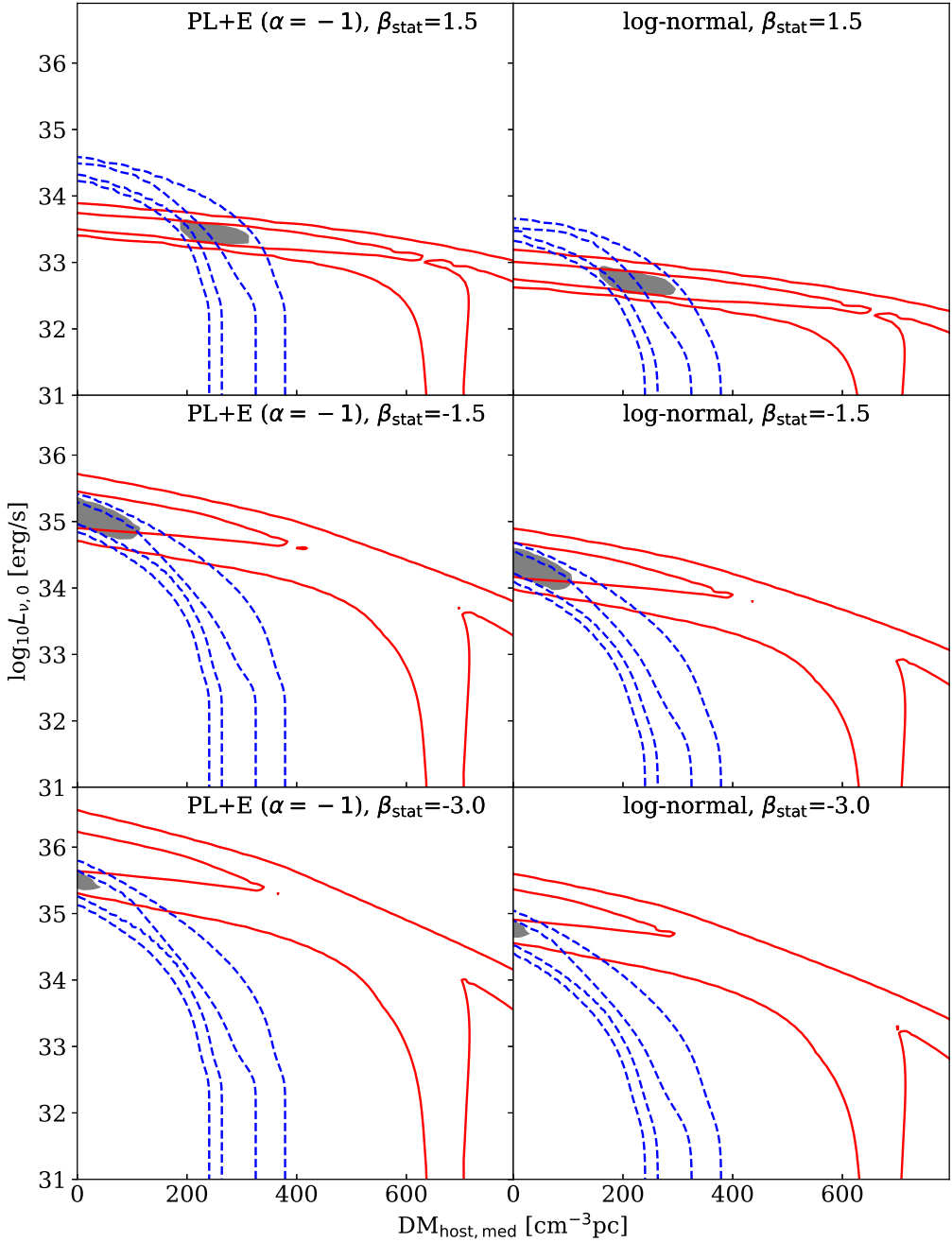} 
\caption{
Same as Figure~\ref{fig:2Dfit_noK}, 
but with $\betast = 1.5$, -1.5, and -3.0 
(top, middle, and bottom panels). 
$\Pks$ likelihood for the baseline PL+E 
and log-normal models of LF are shown 
in the left and right panels, respectively.  
\label{fig:2Dfit_Kcorr} 
}
\end{figure*}

In the previous section, we have assumed that $K$-correction does not affect 
observed flux density of an FRB [$\kappa_\nu(z) = 1$, or $\betast = 0$]. 
In this section, we examine how the results of the $\DMex$ distribution 
fitting changes when $\betast$ is changed. 
The acceptable ranges of the parameters with $\betast = 1.5, -1.5$, and $-3.0$ 
are shown as contours of $\Pks$ in Figure~\ref{fig:2Dfit_Kcorr} 
for the baseline PL+E ($\alpha = -1$) and log-normal models of LF. 
Here the characteristic luminosity density $\Lo$ 
is determined at the emitting frequency observed at $z = 0$ 
($\sim 1.3$ GHz in the case of the Parkes telescope and ASKAP), 
and the characteristic luminosity in other emitting frequency follows $\propto \nu^\betast$. 
The results with the other LF models are not qualitatively different. 

\begin{figure}
\plotone{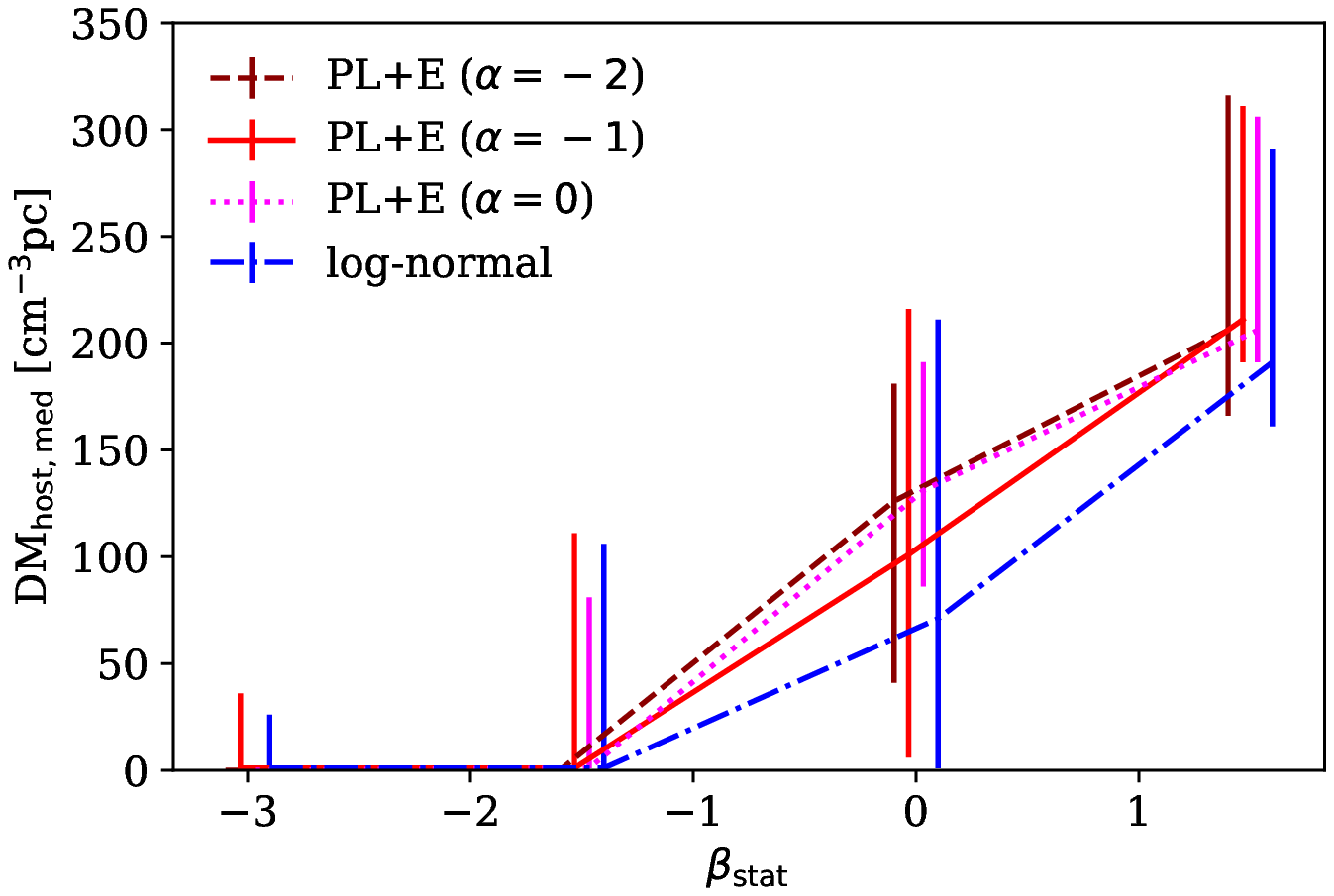}
\caption{
Possible range of $\DMhostmed$ that can simultaneously reproduce 
the observed DM distributions of the Parkes and ASKAP samples 
($P_{\rm KS,joint} > 0.1$) as a function of spectral index $\betast$. 
The datapoints connected with dashed, solid, dotted, and dot-dashed lines 
represent the results with the PL+E ($\alpha = -2.0, -1.0$, 0.0), 
and log-normal models of LF, respectively. 
The data points are slightly shifted sideways for visibility. 
\label{fig:DMhost}
}
\end{figure}

The best-fitting $\Lo$ is larger for smaller $\betast$ as naturally expected, 
and the fitting to the Parkes sample is affected more 
from the $K$-correction than that to the ASKAP sample, 
because FRBs in the Parkes sample have larger $\DMex$ 
(i.e., likely at higher-$z$) on average than those in the ASKAP sample. 
These effects make the preferred $\DMhostmed$ larger with larger $\betast$. 
The range of $\DMhostmed$ which can provide $P_{\rm KS,joint} > 0.1$ 
depending on $\Lo$ is shown in Figure~\ref{fig:DMhost} for the four LF models. 
In the cases of $\betast \leq -1.5$, there is no parameter range 
that reproduce the observed $\DMex$ distributions 
of the Parkes and ASKAP samples at a same time with the steep PL+E LF model. 
On the other hand, $\DMhostmed \gtrsim 200$ cm$^{-3}$pc 
is preferred when $\betast = 1.5$, suggesting existence 
of other significant DM components than ISM and halo gas in FRB host galaxies. 


\section{Distributions of flux density and fluence} 
\label{sec:flux} 

We have seen that preferred amount of $\DMhostmed$ is dependent on $\betast$. 
In this section, we investigate $S_\nu$ (and $F_\nu$) distribution, 
so-called log$N$--log$S$ distribution, of FRBs to constrain $\betast$. 
It is widely known that $S_\nu$ and $F_\nu$ of a population of light sources 
follow a power-law distribution $N(> \mathcal{F}) \propto \mathcal{F}^\gamma$ 
with index $\gamma = -1.5$ ($\mathcal{F} = S_\nu$ or $F_\nu$) 
when the light sources are homogeneously distributed in a Euclidean space, 
while cosmological effects can modify those distributions. 
As discussed in N18, the cosmological effects modify  
distributions of $S_\nu$ and $F_\nu$ differently. 
Hence we investigate both $S_\nu$ and $F_\nu$ distributions in this study. 

\begin{figure*}
\plotone{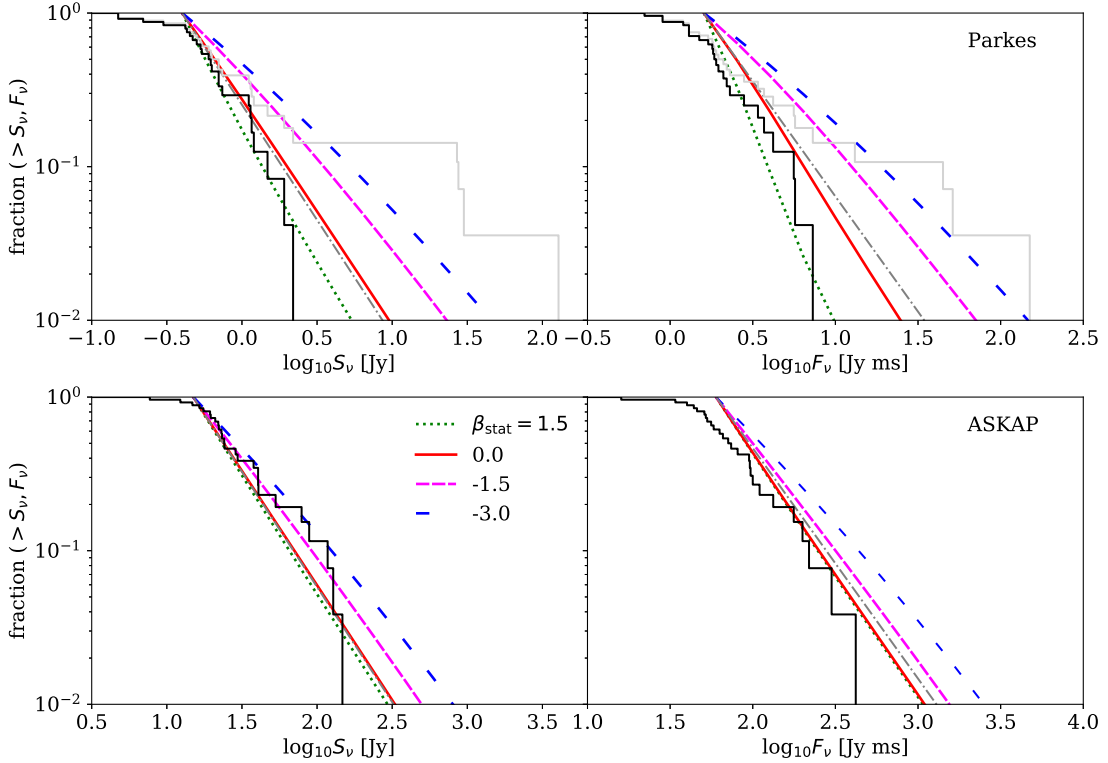} 
\caption{
Left panels: the cumulative distribution of $S_\nu$. 
The results with $\betast=1.5$, 0.0, -1.5, and -3.0 
are shown with dotted, solid, short-dashed, and long-dashed lines, respectively. 
The dot-dashed line indicates the power-law with index $\gamma = -1.5$, 
which is expected for objects homogeneously distributed in a Euclidean space. 
The black histograms show the distribution of the observed samples analyzed in this study, 
The gray histogram in the upper panel shows the distribution 
of the Parkes sample including the four outlier events 
that are excluded from our analysis (see Section~\ref{sec:sample}). 
Right panels: same as the left panels but for $F_\nu$. 
The upper and lower panels show the distributions 
for the Parkes and ASKAP samples, respectively. 
\label{fig:logNlogF}
}
\end{figure*}

We compute $S_\nu$ and $F_\nu$ distributions of model FRBs 
for each combination of a LF model and $\betast$ assuming the $\Lo$ and $\DMhostmed$ 
that provide the highest $P_{\rm KS,joint}$ in the $\DMex$ distribution fitting. 
The computations are done for both the Parkes 
and ASKAP samples (i.e., for the different detection thresholds). 
In Figure~\ref{fig:logNlogF}, we show 
the distributions of $S_\nu$ and $F_\nu$ with the baseline PL+E LF model. 
The results with the other LF models are not significantly different, 
except that we have not computed the distributions 
with the steep PL+E model for $\betast \leq -1.5$, 
for which the $\DMex$ distributions of the two observed FRB samples 
cannot be reproduced at a same time with any set of $\Lo$ and $\DMhostmed$. 

$S_\nu$ and $F_\nu$ of the observed samples are 
also plotted together in Figure~\ref{fig:logNlogF}. 
It is notable that the observed cumulative distributions 
of $S_\nu$ and $F_\nu$ flatten in their faint-end, 
due to the detection incompleteness. 
However, the distribution of $S_\nu$ is more sharply cut 
in the faint-end than the distribution of $F_\nu$ 
in both of the Parkes and ASKAP samples, 
with smaller number of FRBs in the $S_\nu$ ($F_\nu$) range 
affected by the incompleteness effect, 
supporting the hypothesis that $S_\nu$ is a good proxy for S/N.

To compare the predicted $S_\nu$ and $F_\nu$ distributions with observations, 
we compute power-law indices ($\gamma$) of the log$N$-log$S$ distributions that explain 
the observed samples using the maximum likelihood method \citep{Crawford:1970a}. 
To examine effects of the observational incompleteness of faint events to the index $\gamma$, 
we compute $\gamma$ for a subsample of observed FRBs above a threshold $S_\nu$ ($F_\nu$),  
and show how the resulting $\gamma$ vaies as the threshold changes, 
as has been done in \citet{Macquart:2018a} and \citet{Bhandari:2018a}. 

\begin{figure*}
\gridline{\fig{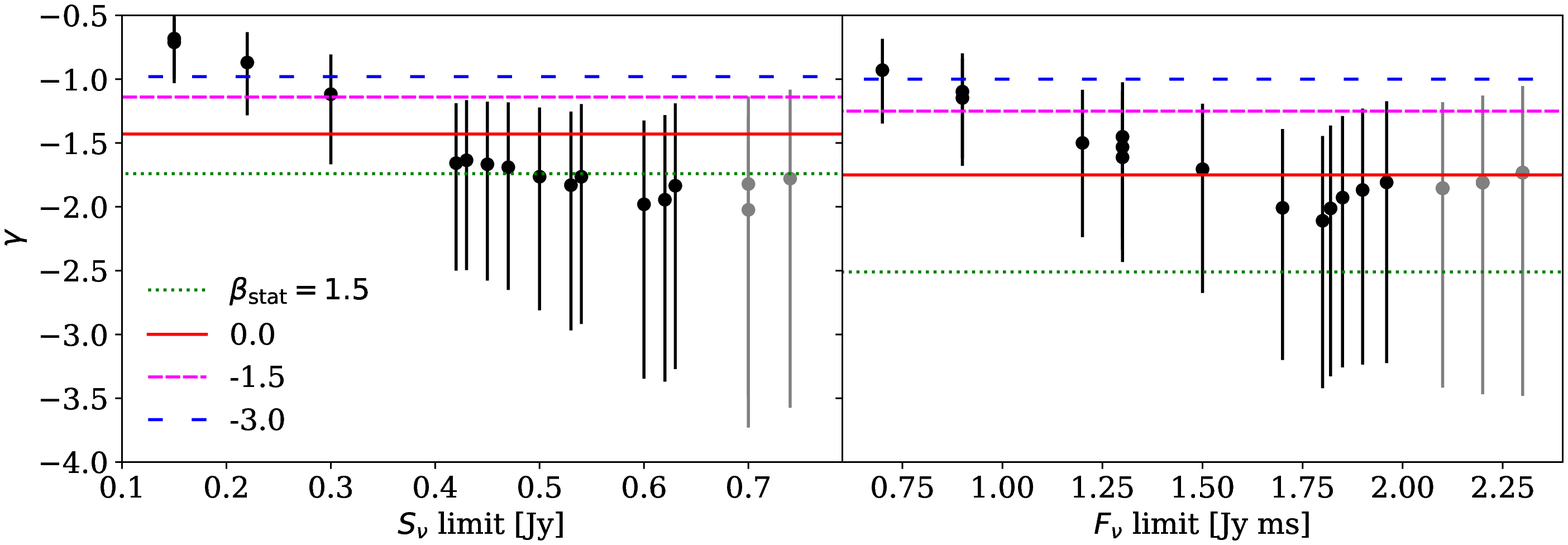}{0.8\textwidth}{(a) Parkes}}
\gridline{\fig{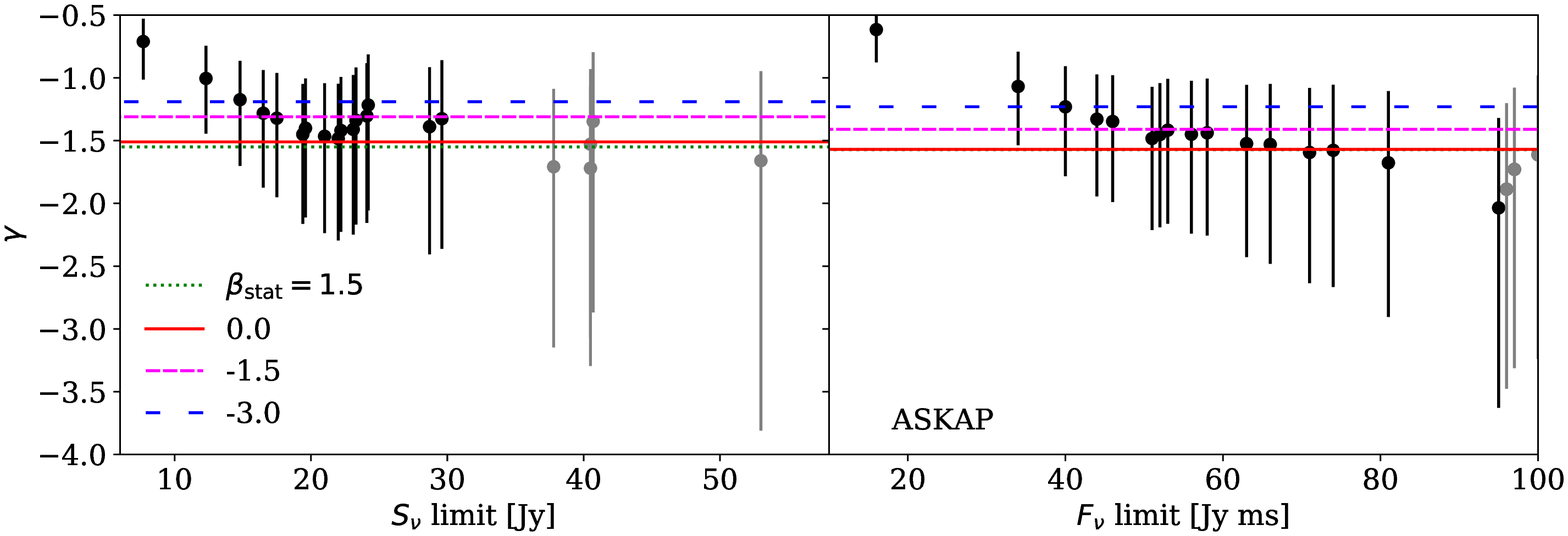}{0.8\textwidth}{(b) ASKAP}}
\caption{
Upper pnaels: 
The power-low index $\gamma$ of the log$N$-log$S$ distribution of the Parkes samples, 
as derived by the maximum likelihood method \citep{Crawford:1970a} 
as a function of the minimum $S_\nu$ ($F_\nu$) 
in the subsample used to derive $\gamma$. 
The errorbars represent the 90\% confidence interval. 
The horizontal lines (dotted, solid, short-dashed, and long-dashed) 
represent the model predictions ($\betast=1.5$, 0.0, -1.5, and -3.0, respectively). 
The gray datapoints are the results derived using subsamples with $\leq 10$ FRBs. 
The left and right panels show $\gamma$ for the $S_\nu$ and $F_\nu$ distributions, respectively.  
Lower panels: same as the upper panels but for the ASKAP sample. 
\label{fig:logNlogFindex} 
} 
\end{figure*}

The obtained $\gamma$ is shown as a function of the minimum $S_\nu$ ($F_\nu$) 
of the subsample used for the computation in Figure~\ref{fig:logNlogFindex}. 
Here we consider allowed range of $\gamma$, 
which is computed using subsample of 20 FRBs 
from the brighter side of the Parkes and ASKAP samples 
[excluding 4 (6) faintest FRBs from the Parkes (ASKAP) sample], 
as the current observational constraints, 
that is $\gamma=-1.66^{+0.47}_{-0.84}$ ($-1.61^{+0.46}_{-0.82}$) 
for the $S_\nu$ ($F_\nu$) distribution of the Parkes sample, 
and $\gamma=-1.40^{+0.40}_{-0.71}$ ($-1.45^{+0.41}_{-0.74}$) 
for the $S_\nu$ ($F_\nu$) distribution of the ASKAP sample. 
The estimation errors of $\gamma$ are 90\% confidence intervals 
computed using the PDF presented in \citet{Crawford:1970a}. 

$\gamma$ predicted by the FRB models 
are also shown in Figure~\ref{fig:logNlogFindex}. 
$S_\nu$ and $F_\nu$ do not strictly follow a power-law 
distribution when the space is not Euclidean, 
and the predicted $\gamma$ is a averaged value in the range 
that the fraction of FRBs with flux density $> S_\nu$ (fluence $> F_\nu$) is larger than 0.01. 
When $\betast=0$, the predictions of both $S_\nu$ and $F_\nu$ 
distributions are consistent with the Parkes and ASKAP samples. 
In the case of $\betast=1.5$, the predicted $\gamma$ 
for the $F_\nu$ distribution of the Parkes sample 
is not consistent with the observational estimate. 
In the cases of $\betast=-1.5$ and $-3.0$, 
the predicted $\gamma$ for the $S_\nu$ distribution 
of the Parkes sample is disfavored. 
The models for the Parkes sample are more strongly affected 
by the cosmological effects than the models 
for the ASKAP sample due to the wider redshift coverage of the sample. 
For the ASKAP sample, the model predictions are consistent 
with the observations regardless of the $K$-correction models. 

\begin{figure*}
\plottwo{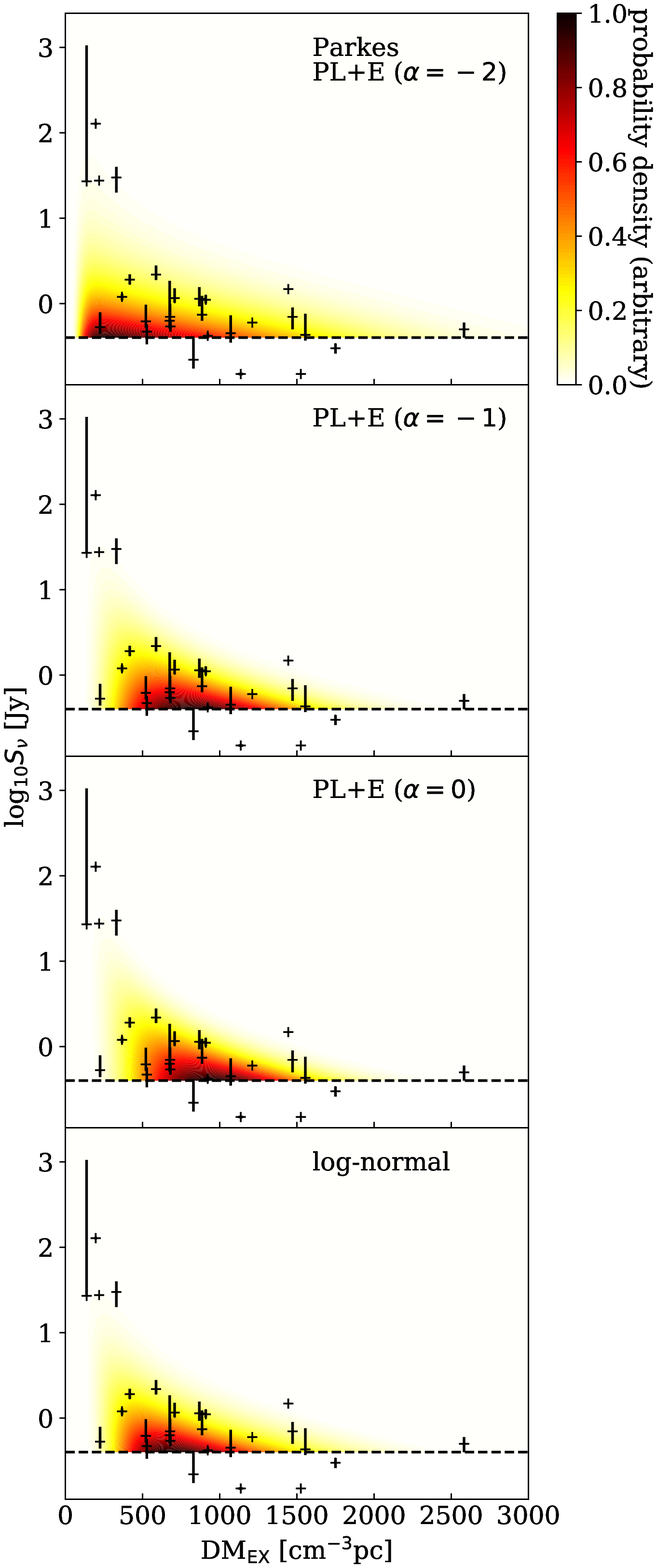}{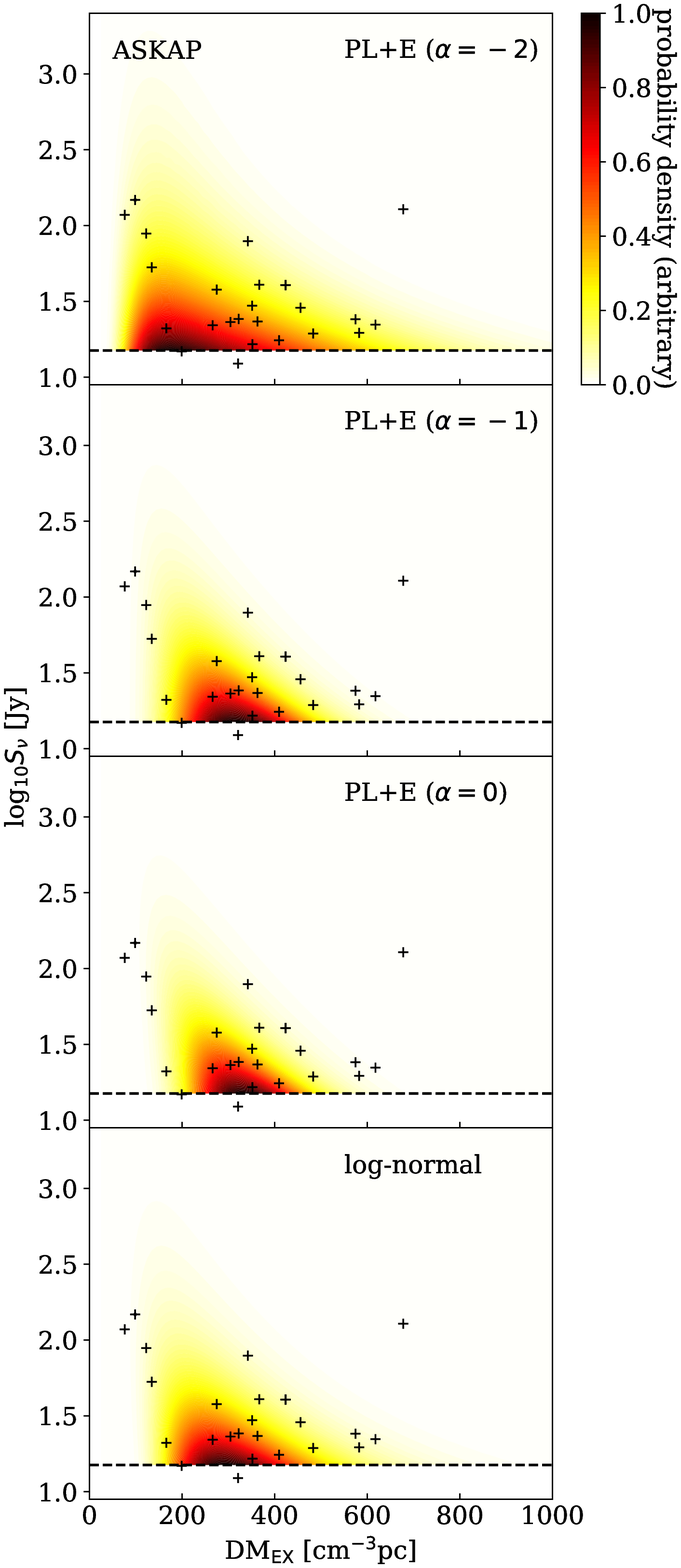} 
\caption{
Left panels: 
probability distributions of FRBs on the parameter plane 
of $\DMex$ vs. $S_\nu$ predicted for the Parkes sample assuming $\betast = 0.0$. 
The results with the PL+E LF models with $\alpha = -2.0, -1.0, 0.0$, 
and the log-normal LF model are shown in the four panels from top to bottom. 
The datapoints are the observational data including the four outlier events 
that are excluded from the analysis ($S_\nu > 10$ Jy, see Section~\ref{sec:sample}). 
The horizontal dashed line in each panel indicates the assumed detection limit. 
Right panels: same as the left panels but for the ASKAP sample. 
\label{fig:DMF2D}
}
\end{figure*}

\begin{figure*}
\gridline{ \fig{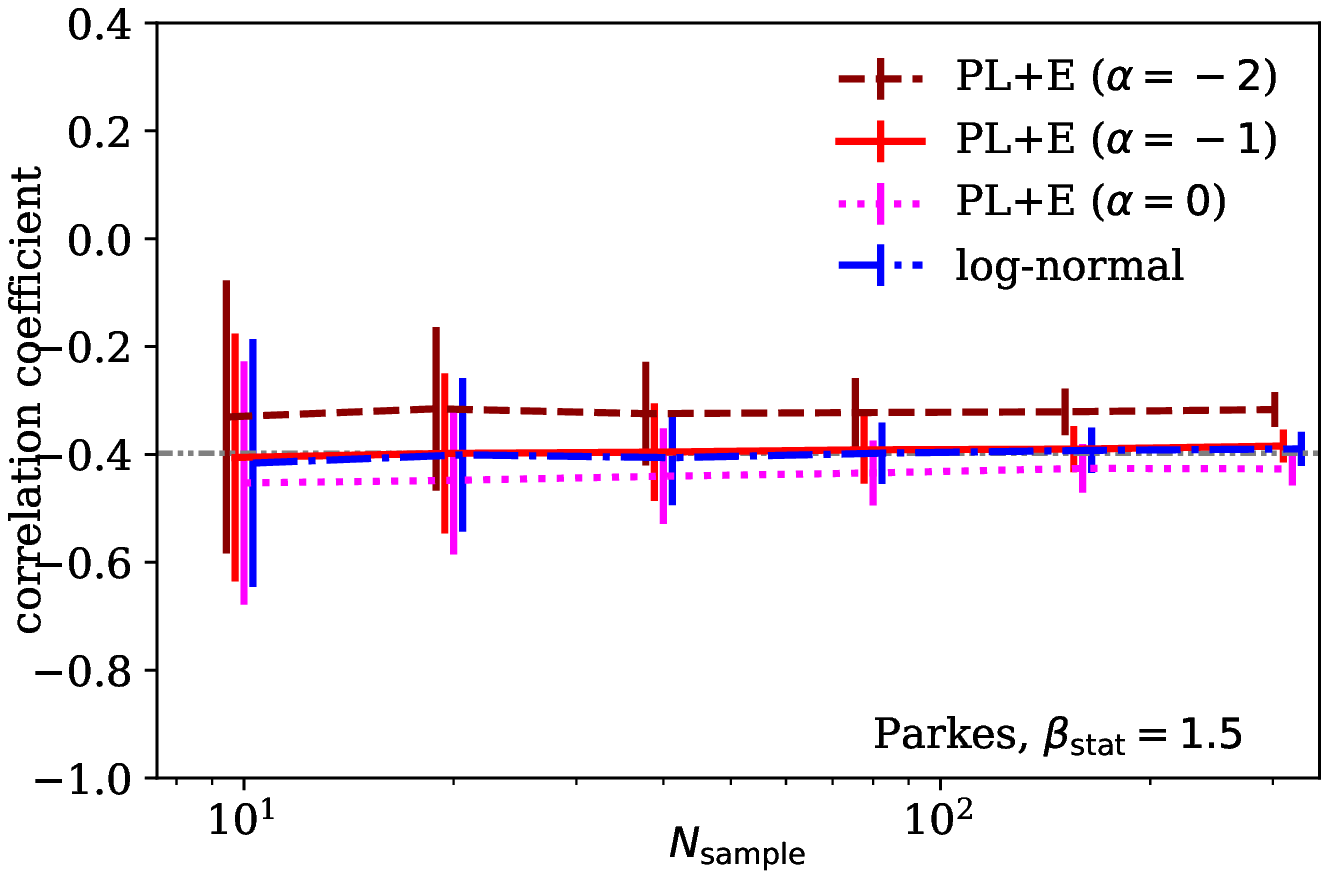}{0.4\textwidth}{}
            \hspace{-40mm}
            \fig{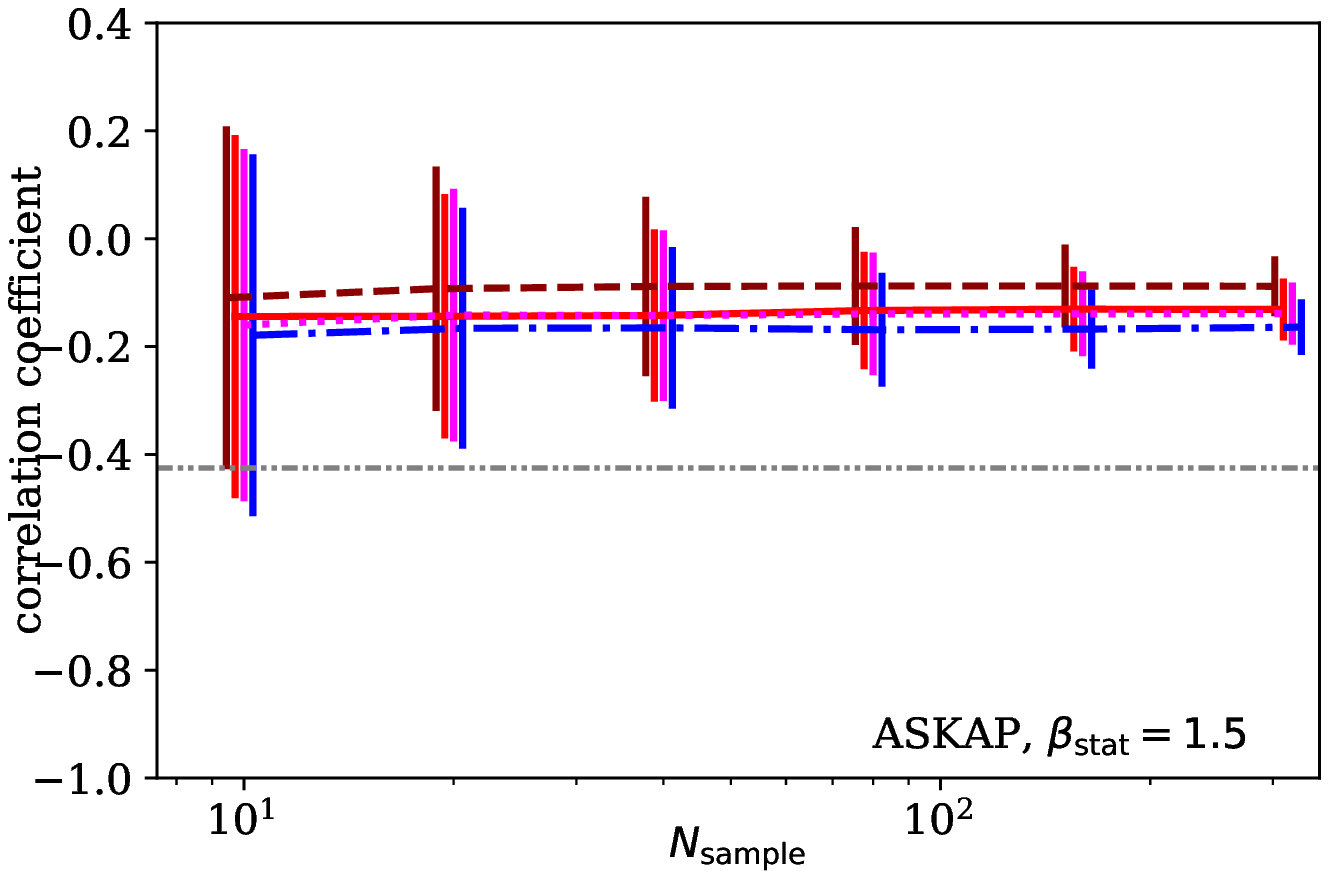}{0.4\textwidth}{} }
\vspace{-10mm}
\gridline{ \fig{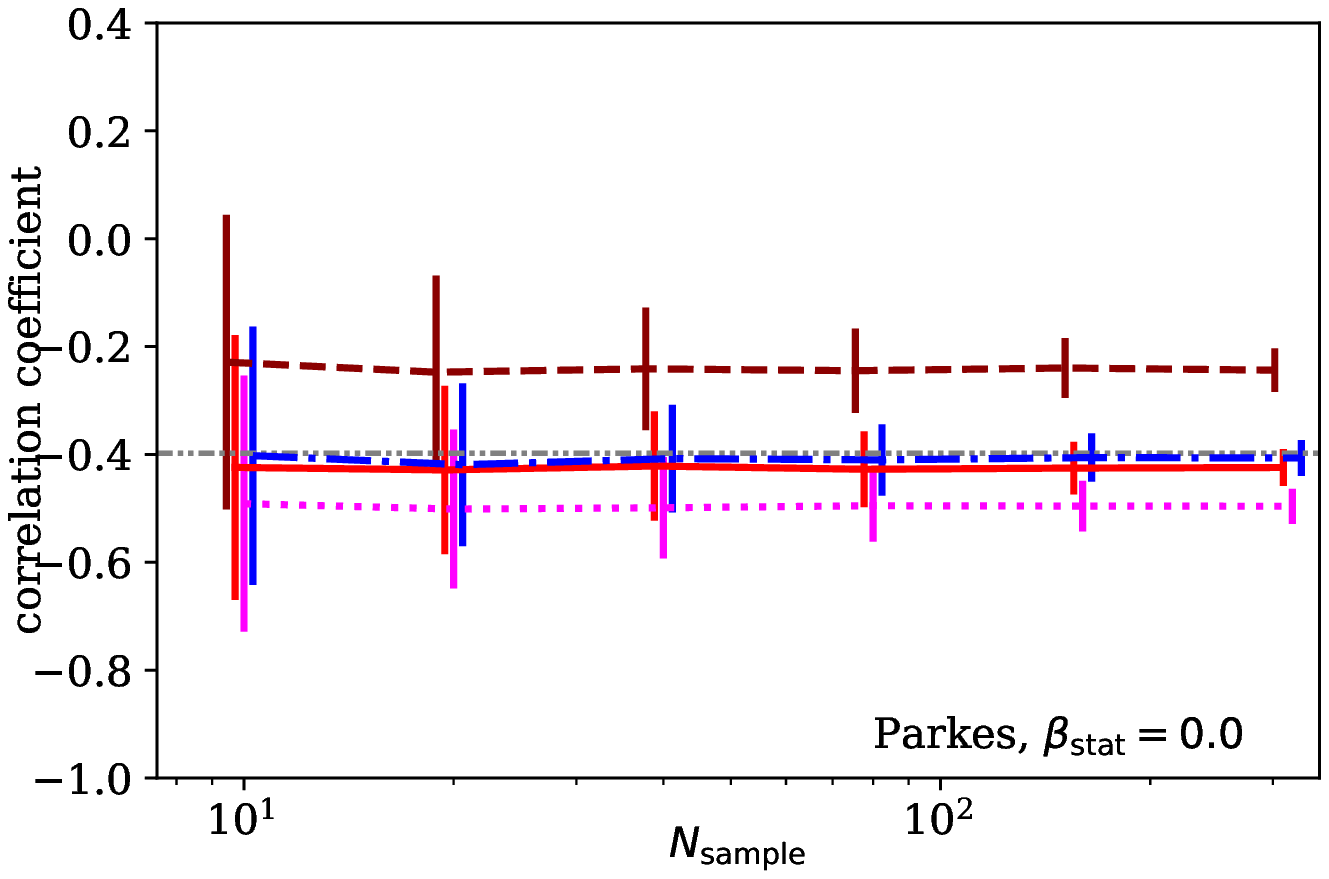}{0.4\textwidth}{}
            \hspace{-40mm}
            \fig{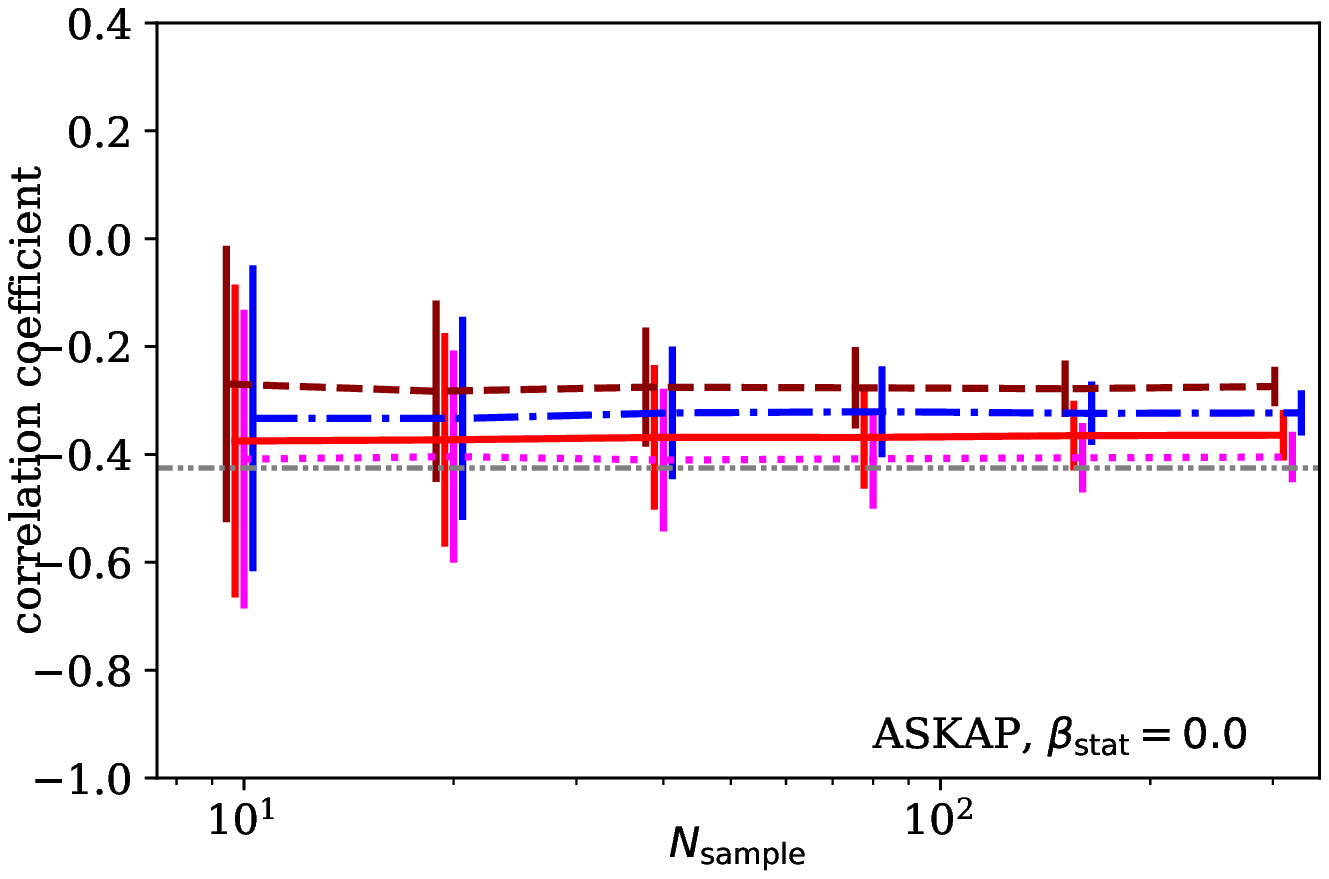}{0.4\textwidth}{} } 
\vspace{-10mm}
\gridline{ \fig{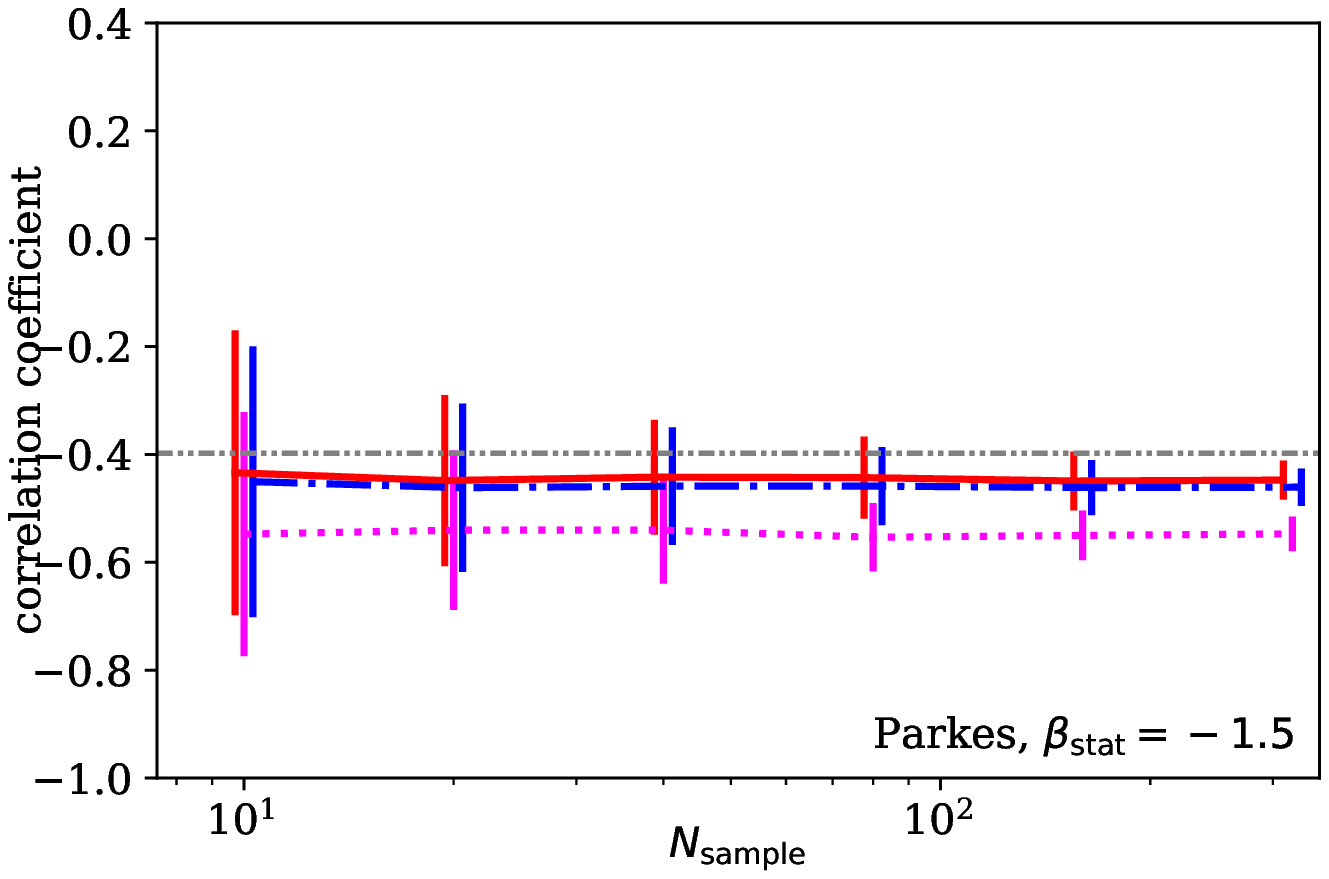}{0.4\textwidth}{}
            \hspace{-40mm}
            \fig{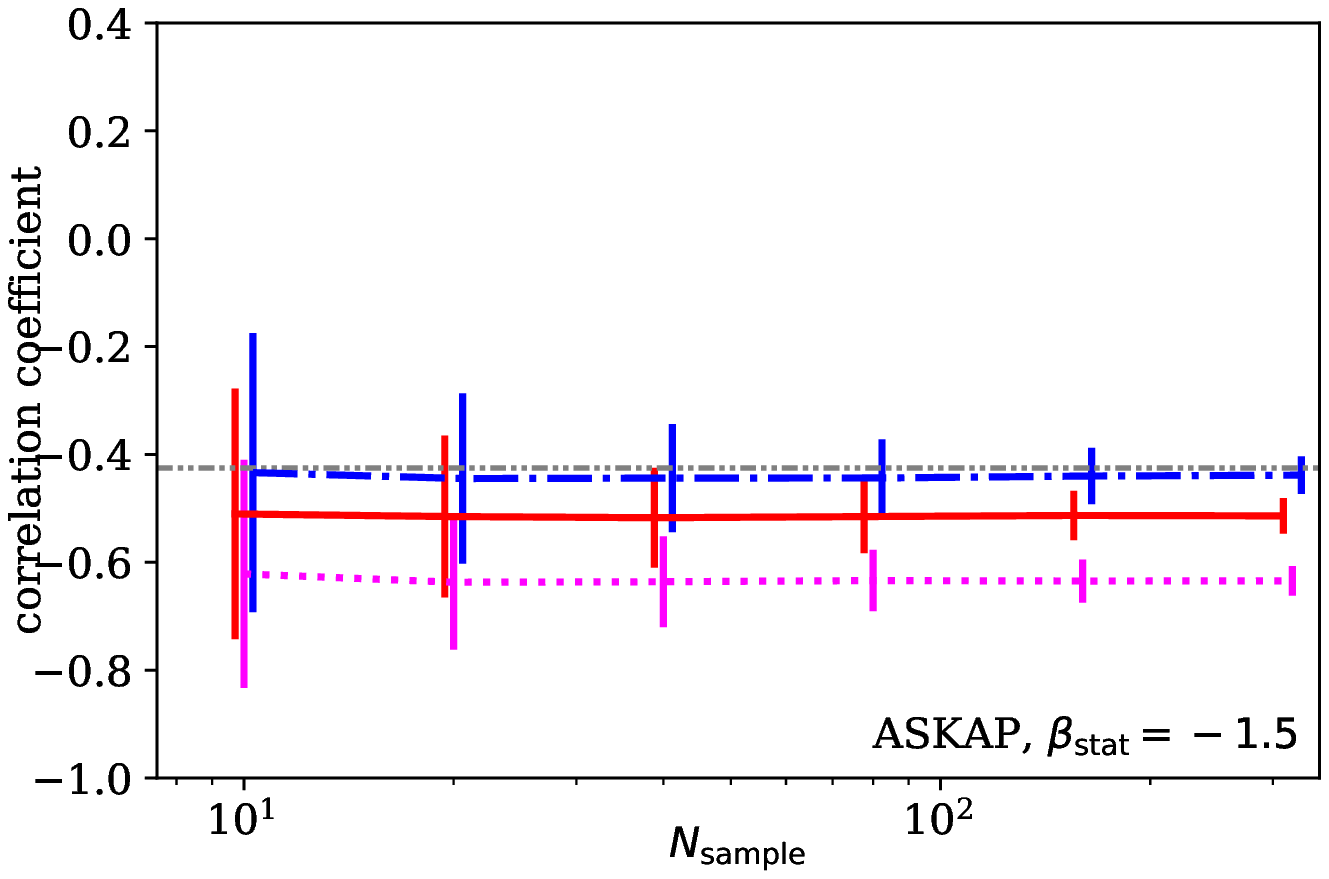}{0.4\textwidth}{} }
\vspace{-10mm}
\gridline{ \fig{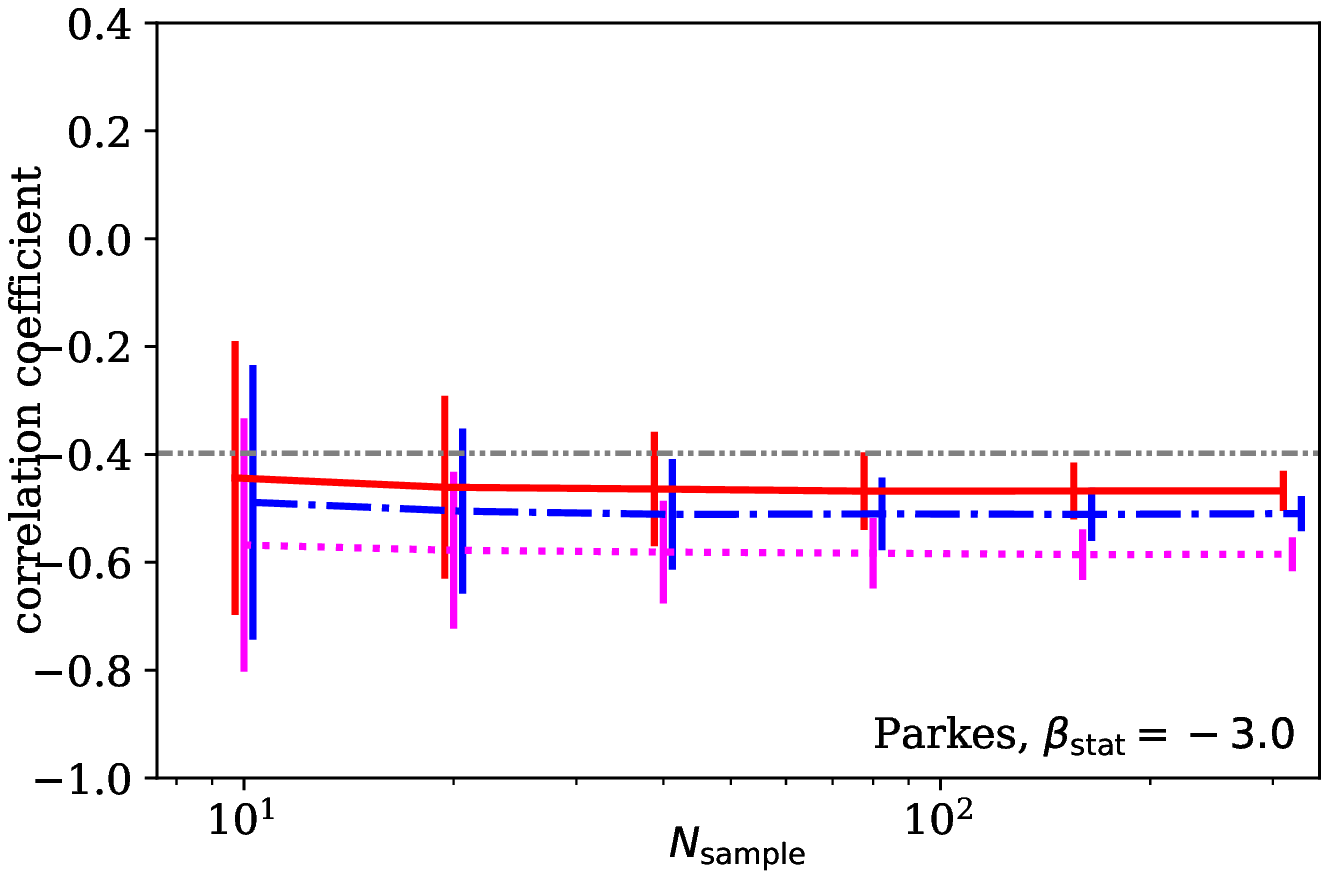}{0.4\textwidth}{}
            \hspace{-40mm}
            \fig{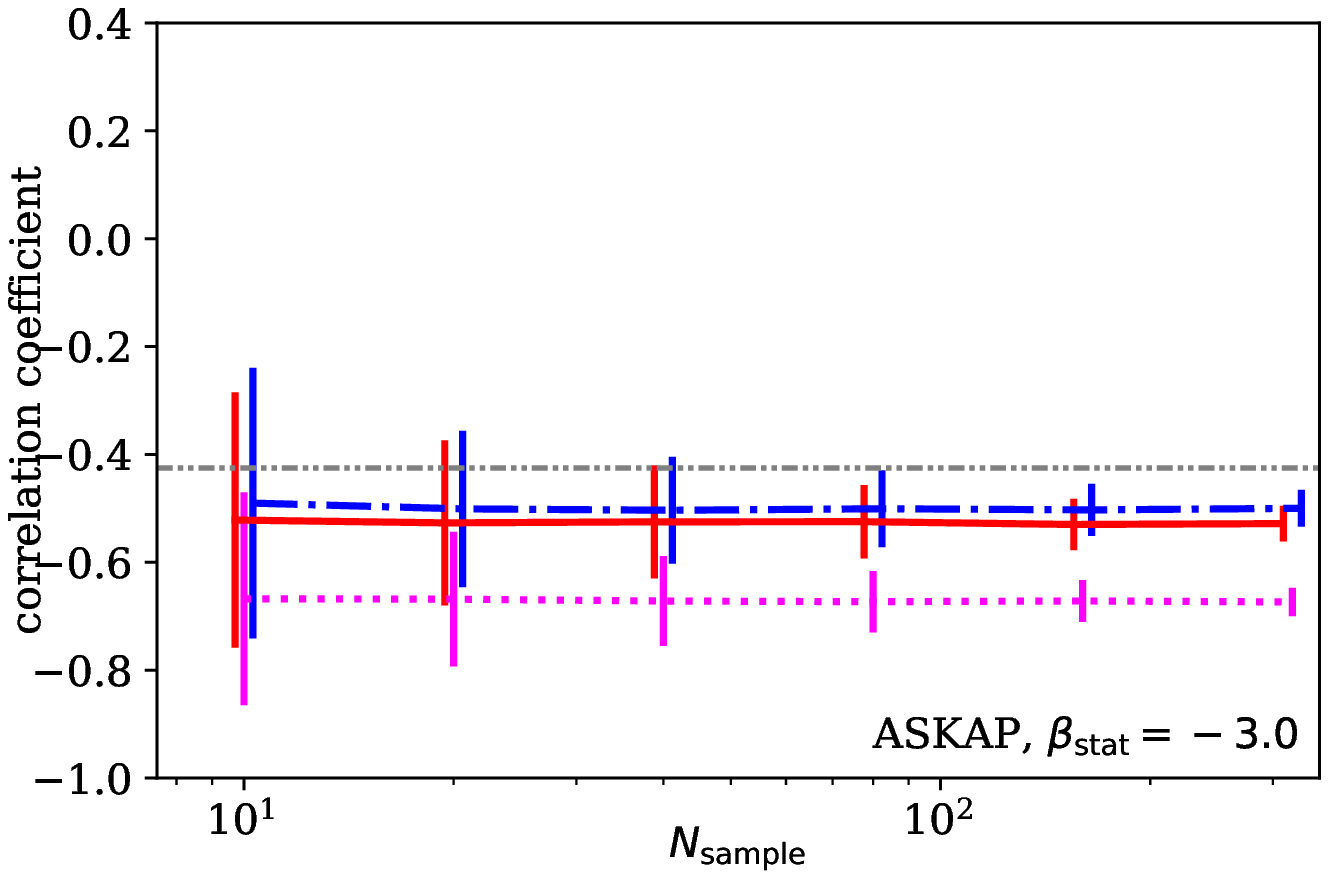}{0.4\textwidth}{} }
\caption{
Left panels: mean and standard deviation 
of the correlation coefficient between $\DMex$ and $S_\nu$ 
generated by the Monte Carlo tests in accordance 
with the distribution functions shown in the left panels 
of figure~\ref{fig:DMF2D} (models for the Parkes sample). 
The random generation of mock sample 
is performed 1000 times for each $N_{\rm sample}$. 
The results with $\betast = 1.5$, 0.0, -1.5, and -3.0 
are shown in the four panels from top to bottom. 
The datapoints connected with dashed, solid, dotted and dot-dashed lines 
show the correlation coefficient distribution with the PL+E LF models 
with $\alpha = -2.0, -1.0, 0.0$, and the log-normal LF model, respectively.
The result with the PL+E LF model with $\alpha = -2.0$ is not shown 
in the two panels from the bottom because the models do not 
have a solution in the $\DMex$ fitting. 
The data points are slightly shifted sideways for visibility. 
The horizontal double-dot-dashed line indicates 
the correlation coefficient between $\DMex$ and $S_\nu$ 
in the observed sample ($N_{\rm sample} = 24$). 
Right panels: same as the left panels but for the ASKAP sample. 
The observed sample has $N_{\rm sample} = 26$. 
\label{fig:CCstat}
}
\end{figure*}

Besides the log$N$--log$S$ distribution, the correlation 
between $\DMex$ and $S_\nu$ can be used as a clue to understand 
the nature of FRBs (\citeauthor{Yang:2017a}~\citeyear{Yang:2017a}; 
N18; \citeauthor{Shannon:2018a}~\citeyear{Shannon:2018a}). 
Here we consider $S_\nu$ rather than $F_\nu$, 
because $S_\nu$ depends more strongly on distance than $F_\nu$ 
and hence the correlation with $\DMex$ becomes more significant. 
Distribution of FRBs on the parameter plane 
of $\DMex$ vs. $S_\nu$ assuming the best-fit parameters 
of the $\DMex$ distribution fitting is shown in Figure~\ref{fig:DMF2D}. 
Following N18, we randomly generate $10^3$ sets of mock samples 
of $\DMex$ and $S_\nu$ with sample size $N_{\rm sample}$ each 
in accordance with the model distributions, 
and compute probability distribution of 
the correlation coefficient between $\DMex$ and $S_\nu$. 
In figure~\ref{fig:CCstat}, we show the mean and the standard deviation 
of the correlation coefficient distributions as functions of $N_{\rm sample}$. 

Among the PL+E LF models, models with steeper faint-end 
shows weaker correlation between $\DMex$ and $S_\nu$, 
because more events are detected near the detection limit regardless of $\DMex$. 
When $\betast = 1.5$, all the LF models predict weak correlation 
between $\DMex$ and $S_\nu$ for the ASKAP sample, 
because $\DMhostmed \sim 200$ cm$^{-3}$pc 
that is determined by the $\DMex$ distribution fitting occupies 
significant fraction of typical $\DMex \sim 350$ cm$^{-3}$pc of the ASKAP sample. 
However, the difference of the predictions 
between the models is difficult to distinguish 
with the current sample size (within $2\sigma$). 

It should be also noted that a $\DMex$--$S_\nu$ correlation 
can be artificially produced by dispersion smearing 
which broadens pulse width (decreases $S_\nu$) more for FRBs with larger DM. 
However, as mentioned in N18, 
pulse width of the FRBs in the Parkes sample 
is not correlated with their $\DMex$ (correlation coefficient $= 0.11$), 
suggesting that the $\DMex$--$S_\nu$ correlation 
is not primarily produced by dispersion smearing effect. 
On the other hand, pulse width of the FRBs in the ASKAP sample 
is correlated with $\DMex$ (correlation coefficient $= 0.49$), 
and hence it is possible that the $\DMex$--$S_\nu$ 
correlation in the ASKAP sample is artificial. 
This is possibly due to the higher spectral resolution 
of the the Parkes sample \citep[$\Delta\nu\sim0.4$~MHz,][]{Crawford:2016a} 
than the ASKAP sample \citep[$\Delta\nu\sim1$~MHz,][]{Bannister:2017a}. 

\section{Conclusions} 
\label{sec:conc} 

We have computed $\DMex$ distribution, 
log$N$-log$S$ distribution, and $\DMex$--$S_\nu$ correlation 
using an analytic model of FRB population that includes 
a variation of LF and $K$-correction models. 
FRB models that fulfill the following conditions 
are favored to reproduce the statistical properties 
of the Parkes and ASKAP samples of FRBs simultaneously, 
\begin{enumerate}
\item DM components that are directly associated 
with FRB progenitors (CSM) are typically small ($< 80$ cm$^{-3}$pc), 
\item LF of FRBs has a bright-end cutoff 
at log$_{10}L_\nu$ [erg s$^{-1}$Hz$^{-1}$] $\sim$ 34, 
\item statistical effect of $K$-correction on 
observed flux density (or fluence) is smaller than a factor of 5 
in the redshift range of $z \lesssim 2$ (i.e. $|\betast| < 1.5$). 
In other words, the typical luminosity density 
of FRBs does not largely changed within the range 
of emitting frequency $\nu_{\rm rest} \sim$ 1--4 GHz. 
\end{enumerate}
although the statistical significance of the constraints are still low ($\sim 90$\%) 
and larger sample of observed FRBs are necessary to obtain robust conclusions. 

The conditions 1 \& 2 are required to reproduce the $\DMex$ distributions 
of the Parkes and ASKAP samples simultaneously (\S~\ref{sec:2Dfitting}). 
Although a larger DM component of $\gtrsim 100$ cm$^{-3}$pc
can be associated with FRB progenitors 
if there is a strong negative $K$-correction effect ($\betast \sim 1.5$).  
However, models  with $|\betast| > 1.5$ 
are disfavored by the observed log$N$--log$S$ distribution 
of the Parkes sample (condition 3, \S~\ref{sec:flux}). 

The constraint on $\DMhostmed$ obtained by our analysis 
indicates that major part of $\DMex$ of an FRB indeed 
arise from the IGM and is a good indicator of distance, 
that will help us to understand FRBs in terms 
of their distance distribution and energetics. 
Together with more robust test of the results 
shown here by larger sample size 
and redshift measurements of FRBs, 
FRBs will also provide us with an unprecedented 
opportunity to study the IGM observationally.  

\acknowledgments
This research has been supported 
by JSPS KAKENHI Grant Number JP17K14255, and JP20H01942. 

\bibliography{reference_list}

\begin{thebibliography}{}
\expandafter\ifx\csname natexlab\endcsname\relax\def\natexlab#1{#1}\fi
\providecommand{\url}[1]{\href{#1}{#1}}
\providecommand{\dodoi}[1]{doi:~\href{http://doi.org/#1}{\nolinkurl{#1}}}
\providecommand{\doeprint}[1]{\href{http://ascl.net/#1}{\nolinkurl{http://ascl.net/#1}}}
\providecommand{\doarXiv}[1]{\href{https://arxiv.org/abs/#1}{\nolinkurl{https://arxiv.org/abs/#1}}}

\bibitem[{{Bannister} {et~al.}(2017){Bannister}, {Shannon}, {Macquart},
  {Flynn}, {Edwards}, {O'Neill}, {Os{\l}owski}, {Bailes}, {Zackay}, {Clarke},
  {D'Addario}, {Dodson}, {Hall}, {Jameson}, {Jones}, {Navarro}, {Trinh},
  {Allison}, {Anderson}, {Bell}, {Chippendale}, {Collier}, {Heald}, {Heywood},
  {Hotan}, {Lee-Waddell}, {Madrid}, {Marvil}, {McConnell}, {Popping},
  {Voronkov}, {Whiting}, {Allen}, {Bock}, {Brodrick}, {Cooray}, {DeBoer},
  {Diamond}, {Ekers}, {Gough}, {Hampson}, {Harvey-Smith}, {Hay}, {Hayman},
  {Jackson}, {Johnston}, {Koribalski}, {McClure-Griffiths}, {Mirtschin}, {Ng},
  {Norris}, {Pearce}, {Phillips}, {Roxby}, {Troup}, \&
  {Westmeier}}]{Bannister:2017a}
{Bannister}, K.~W., {Shannon}, R.~M., {Macquart}, J.~P., {et~al.} 2017, \apjl,
  841, L12, \dodoi{10.3847/2041-8213/aa71ff}

\bibitem[{{Bannister} {et~al.}(2019){Bannister}, {Deller}, {Phillips},
  {Macquart}, {Prochaska}, {Tejos}, {Ryder}, {Sadler}, {Shannon}, {Simha},
  {Day}, {McQuinn}, {North-Hickey}, {Bhandari}, {Arcus}, {Bennert}, {Burchett},
  {Bouwhuis}, {Dodson}, {Ekers}, {Farah}, {Flynn}, {James}, {Kerr}, {Lenc},
  {Mahony}, {O'Meara}, {Os{\l}owski}, {Qiu}, {Treu}, {U}, {Bateman}, {Bock},
  {Bolton}, {Brown}, {Bunton}, {Chippendale}, {Cooray}, {Cornwell}, {Gupta},
  {Hayman}, {Kesteven}, {Koribalski}, {MacLeod}, {McClure-Griffiths},
  {Neuhold}, {Norris}, {Pilawa}, {Qiao}, {Reynolds}, {Roxby}, {Shimwell},
  {Voronkov}, \& {Wilson}}]{Bannister:2019a}
{Bannister}, K.~W., {Deller}, A.~T., {Phillips}, C., {et~al.} 2019, Science,
  365, 565, \dodoi{10.1126/science.aaw5903}

\bibitem[{{Bhandari} {et~al.}(2018){Bhandari}, {Keane}, {Barr}, {Jameson},
  {Petroff}, {Johnston}, {Bailes}, {Bhat}, {Burgay}, {Burke-Spolaor}, {Caleb},
  {Eatough}, {Flynn}, {Green}, {Jankowski}, {Kramer}, {Krishnan}, {Morello},
  {Possenti}, {Stappers}, {Tiburzi}, {van Straten}, {Andreoni}, {Butterley},
  {Chand ra}, {Cooke}, {Corongiu}, {Coward}, {Dhillon}, {Dodson}, {Hardy},
  {Howell}, {Jaroenjittichai}, {Klotz}, {Littlefair}, {Marsh}, {Mickaliger},
  {Muxlow}, {Perrodin}, {Pritchard}, {Sawangwit}, {Terai}, {Tominaga}, {Torne},
  {Totani}, {Trois}, {Turpin}, {Niino}, {Wilson}, {Albert}, {Andr{\'e}},
  {Anghinolfi}, {Anton}, {Ardid}, {Aubert}, {Avgitas}, {Baret},
  {Barrios-Mart{\'\i}}, {Basa}, {Belhorma}, {Bertin}, {Biagi}, {Bormuth},
  {Bourret}, {Bouwhuis}, {Br{\^a}nza{\textcommabelow s}}, {Bruijn}, {Brunner},
  {Busto}, {Capone}, {Caramete}, {Carr}, {Celli}, {Moursli}, {Chiarusi},
  {Circella}, {Coelho}, {Coleiro}, {Coniglione}, {Costantini}, {Coyle},
  {Creusot}, {D{\'\i}az}, {Deschamps}, {De Bonis}, {Distefano}, {Palma},
  {Domi}, {Donzaud}, {Dornic}, {Drouhin}, {Eberl}, {Bojaddaini}, {Khayati},
  {Els{\"a}sser}, {Enzenh{\"o}fer}, {Ettahiri}, {Fassi}, {Felis}, {Fusco},
  {Gay}, {Giordano}, {Glotin}, {Gregoire}, {Gracia-Ruiz}, {Graf}, {Hallmann},
  {van Haren}, {Heijboer}, {Hello}, {Hern{\'a}ndez-Rey}, {H{\"o}{\ss}l},
  {Hofest{\"a}dt}, {Hugon}, {Illuminati}, {James}, {de Jong}, {Jongen},
  {Kadler}, {Kalekin}, {Katz}, {Kie{\ss}ling}, {Kouchner}, {Kreter},
  {Kreykenbohm}, {Kulikovskiy}, {Lachaud}, {Lahmann}, {Lef{\`e}vre}, {Leonora},
  {Loucatos}, {Marcelin}, {Margiotta}, {Marinelli}, {Mart{\'\i}nez-Mora},
  {Mele}, {Melis}, {Michael}, {Migliozzi}, {Moussa}, {Navas}, {Nezri},
  {Organokov}, {P{\v{a}}v{\v{a}}la{\textcommabelow s}}, {Pellegrino},
  {Perrina}, {Piattelli}, {Popa}, {Pradier}, {Quinn}, {Racca}, {Riccobene},
  {S{\'a}nchez-Losa}, {Salda{\~n}a}, {Salvadori}, {Samtleben}, {Sanguineti},
  {Sapienza}, {Sch{\"u}ssler}, {Sieger}, {Spurio}, {Stolarczyk}, {Taiuti},
  {Tayalati}, {Trovato}, {Turpin}, {T{\"o}nnis}, {Vallage}, {Van Elewyck},
  {Versari}, {Vivolo}, {Vizzocca}, {Wilms}, {Zornoza}, \&
  {Z{\'u}{\~n}iga}}]{Bhandari:2018a}
{Bhandari}, S., {Keane}, E.~F., {Barr}, E.~D., {et~al.} 2018, \mnras, 475,
  1427, \dodoi{10.1093/mnras/stx3074}

\bibitem[{{Bochenek} {et~al.}(2020){Bochenek}, {Ravi}, {Belov}, {Hallinan},
  {Kocz}, {Kulkarni}, \& {McKenna}}]{Bochenek:2020a}
{Bochenek}, C.~D., {Ravi}, V., {Belov}, K.~V., {et~al.} 2020, arXiv e-prints,
  arXiv:2005.10828.
\newblock \doarXiv{2005.10828}

\bibitem[{{Caleb} {et~al.}(2016){Caleb}, {Flynn}, {Bailes}, {Barr}, {Hunstead},
  {Keane}, {Ravi}, \& {van Straten}}]{Caleb:2016a}
{Caleb}, M., {Flynn}, C., {Bailes}, M., {et~al.} 2016, \mnras, 458, 708,
  \dodoi{10.1093/mnras/stw175}

\bibitem[{{Cordes} \& {Lazio}(2002)}]{Cordes:2002a}
{Cordes}, J.~M., \& {Lazio}, T.~J.~W. 2002, ArXiv Astrophysics e-prints

\bibitem[{{Cordes} \& {Wasserman}(2016)}]{Cordes:2016b}
{Cordes}, J.~M., \& {Wasserman}, I. 2016, \mnras, 457, 232,
  \dodoi{10.1093/mnras/stv2948}

\bibitem[{{Cordes} {et~al.}(2017){Cordes}, {Wasserman}, {Hessels}, {Lazio},
  {Chatterjee}, \& {Wharton}}]{Cordes:2017a}
{Cordes}, J.~M., {Wasserman}, I., {Hessels}, J.~W.~T., {et~al.} 2017, \apj,
  842, 35, \dodoi{10.3847/1538-4357/aa74da}

\bibitem[{{Cordes} {et~al.}(2016){Cordes}, {Wharton}, {Spitler}, {Chatterjee},
  \& {Wasserman}}]{Cordes:2016a}
{Cordes}, J.~M., {Wharton}, R.~S., {Spitler}, L.~G., {Chatterjee}, S., \&
  {Wasserman}, I. 2016, ArXiv e-prints.
\newblock \doarXiv{1605.05890}

\bibitem[{{Crawford} {et~al.}(1970){Crawford}, {Jauncey}, \&
  {Murdoch}}]{Crawford:1970a}
{Crawford}, D.~F., {Jauncey}, D.~L., \& {Murdoch}, H.~S. 1970, \apj, 162, 405,
  \dodoi{10.1086/150672}

\bibitem[{{Crawford} {et~al.}(2016){Crawford}, {Rane}, {Tran}, {Rolph},
  {Lorimer}, \& {Ridley}}]{Crawford:2016a}
{Crawford}, F., {Rane}, A., {Tran}, L., {et~al.} 2016, \mnras, 460, 3370,
  \dodoi{10.1093/mnras/stw1233}

\bibitem[{{DeLaunay} {et~al.}(2016){DeLaunay}, {Fox}, {Murase},
  {M{\'e}sz{\'a}ros}, {Keivani}, {Messick}, {Mostaf{\'a}}, {Oikonomou},
  {Te{\v{s}}i{\'c}}, \& {Turley}}]{DeLaunay:2016a}
{DeLaunay}, J.~J., {Fox}, D.~B., {Murase}, K., {et~al.} 2016, \apjl, 832, L1,
  \dodoi{10.3847/2041-8205/832/1/L1}

\bibitem[{{Dolag} {et~al.}(2015){Dolag}, {Gaensler}, {Beck}, \&
  {Beck}}]{Dolag:2015a}
{Dolag}, K., {Gaensler}, B.~M., {Beck}, A.~M., \& {Beck}, M.~C. 2015, \mnras,
  451, 4277, \dodoi{10.1093/mnras/stv1190}

\bibitem[{{Falcke} \& {Rezzolla}(2014)}]{Falcke:2014a}
{Falcke}, H., \& {Rezzolla}, L. 2014, \aap, 562, A137,
  \dodoi{10.1051/0004-6361/201321996}

\bibitem[{{Hassall} {et~al.}(2013){Hassall}, {Keane}, \&
  {Fender}}]{Hassall:2013a}
{Hassall}, T.~E., {Keane}, E.~F., \& {Fender}, R.~P. 2013, \mnras, 436, 371,
  \dodoi{10.1093/mnras/stt1598}

\bibitem[{{Inoue}(2004)}]{Inoue:2004a}
{Inoue}, S. 2004, \mnras, 348, 999, \dodoi{10.1111/j.1365-2966.2004.07359.x}

\bibitem[{{Ioka}(2003)}]{Ioka:2003a}
{Ioka}, K. 2003, \apjl, 598, L79, \dodoi{10.1086/380598}

\bibitem[{{Kashiyama} {et~al.}(2013){Kashiyama}, {Ioka}, \&
  {M{\'e}sz{\'a}ros}}]{Kashiyama:2013a}
{Kashiyama}, K., {Ioka}, K., \& {M{\'e}sz{\'a}ros}, P. 2013, \apjl, 776, L39,
  \dodoi{10.1088/2041-8205/776/2/L39}

\bibitem[{{Katz}(2016)}]{Katz:2016a}
{Katz}, J.~I. 2016, \apj, 818, 19, \dodoi{10.3847/0004-637X/818/1/19}

\bibitem[{{Keane} \& {Petroff}(2015)}]{Keane:2015a}
{Keane}, E.~F., \& {Petroff}, E. 2015, \mnras, 447, 2852,
  \dodoi{10.1093/mnras/stu2650}

\bibitem[{{Keane} {et~al.}(2012){Keane}, {Stappers}, {Kramer}, \&
  {Lyne}}]{Keane:2012a}
{Keane}, E.~F., {Stappers}, B.~W., {Kramer}, M., \& {Lyne}, A.~G. 2012, \mnras,
  425, L71, \dodoi{10.1111/j.1745-3933.2012.01306.x}

\bibitem[{{Keane} {et~al.}(2016){Keane}, {Johnston}, {Bhandari}, {Barr},
  {Bhat}, {Burgay}, {Caleb}, {Flynn}, {Jameson}, {Kramer}, {Petroff},
  {Possenti}, {van Straten}, {Bailes}, {Burke-Spolaor}, {Eatough}, {Stappers},
  {Totani}, {Honma}, {Furusawa}, {Hattori}, {Morokuma}, {Niino}, {Sugai},
  {Terai}, {Tominaga}, {Yamasaki}, {Yasuda}, {Allen}, {Cooke}, {Jencson},
  {Kasliwal}, {Kaplan}, {Tingay}, {Williams}, {Wayth}, {Chandra}, {Perrodin},
  {Berezina}, {Mickaliger}, \& {Bassa}}]{Keane:2016a}
{Keane}, E.~F., {Johnston}, S., {Bhandari}, S., {et~al.} 2016, \nat, 530, 453,
  \dodoi{10.1038/nature17140}

\bibitem[{{Kokubo} {et~al.}(2017){Kokubo}, {Mitsuda}, {Sugai}, {Ozaki},
  {Minowa}, {Hattori}, {Hayano}, {Matsubayashi}, {Shimono}, {Sako}, \&
  {Doi}}]{Kokubo:2017a}
{Kokubo}, M., {Mitsuda}, K., {Sugai}, H., {et~al.} 2017, \apj, 844, 95,
  \dodoi{10.3847/1538-4357/aa7b2d}

\bibitem[{{Lorimer} {et~al.}(2007){Lorimer}, {Bailes}, {McLaughlin},
  {Narkevic}, \& {Crawford}}]{Lorimer:2007a}
{Lorimer}, D.~R., {Bailes}, M., {McLaughlin}, M.~A., {Narkevic}, D.~J., \&
  {Crawford}, F. 2007, Science, 318, 777, \dodoi{10.1126/science.1147532}

\bibitem[{{Lu} {et~al.}(2020){Lu}, {Kumar}, \& {Zhang}}]{Lu:2020a}
{Lu}, W., {Kumar}, P., \& {Zhang}, B. 2020, arXiv e-prints, arXiv:2005.06736.
\newblock \doarXiv{2005.06736}

\bibitem[{{Lu} \& {Piro}(2019)}]{Lu:2019a}
{Lu}, W., \& {Piro}, A.~L. 2019, \apj, 883, 40,
  \dodoi{10.3847/1538-4357/ab3796}

\bibitem[{{Luo} {et~al.}(2018){Luo}, {Lee}, {Lorimer}, \& {Zhang}}]{Luo:2018a}
{Luo}, R., {Lee}, K., {Lorimer}, D.~R., \& {Zhang}, B. 2018, \mnras, 481, 2320,
  \dodoi{10.1093/mnras/sty2364}

\bibitem[{{Luo} {et~al.}(2020){Luo}, {Men}, {Lee}, {Wang}, {Lorimer}, \&
  {Zhang}}]{Luo:2020a}
{Luo}, R., {Men}, Y., {Lee}, K., {et~al.} 2020, \mnras, 494, 665,
  \dodoi{10.1093/mnras/staa704}

\bibitem[{{Macquart} \& {Ekers}(2018)}]{Macquart:2018a}
{Macquart}, J.-P., \& {Ekers}, R.~D. 2018, \mnras, 474, 1900,
  \dodoi{10.1093/mnras/stx2825}

\bibitem[{{Madau} \& {Dickinson}(2014)}]{Madau:2014a}
{Madau}, P., \& {Dickinson}, M. 2014, \araa, 52, 415,
  \dodoi{10.1146/annurev-astro-081811-125615}

\bibitem[{{Marcote} {et~al.}(2020){Marcote}, {Nimmo}, {Hessels}, {Tendulkar},
  {Bassa}, {Paragi}, {Keimpema}, {Bhardwaj}, {Karuppusamy}, {Kaspi}, {Law},
  {Michilli}, {Aggarwal}, {Andersen}, {Archibald}, {Bandura}, {Bower}, {Boyle},
  {Brar}, {Burke-Spolaor}, {Butler}, {Cassanelli}, {Chawla}, {Demorest},
  {Dobbs}, {Fonseca}, {Giri}, {Good}, {Gourdji}, {Josephy}, {Kirichenko},
  {Kirsten}, {Landecker}, {Lang}, {Lazio}, {Li}, {Lin}, {Linford}, {Masui},
  {Mena-Parra}, {Naidu}, {Ng}, {Patel}, {Pen}, {Pleunis}, {Rafiei-Ravandi},
  {Rahman}, {Renard}, {Scholz}, {Siegel}, {Smith}, {Stairs}, {Vanderlinde}, \&
  {Zwaniga}}]{Marcote:2020a}
{Marcote}, B., {Nimmo}, K., {Hessels}, J.~W.~T., {et~al.} 2020, \nat, 577, 190,
  \dodoi{10.1038/s41586-019-1866-z}

\bibitem[{{Margalit} {et~al.}(2020){Margalit}, {Beniamini}, {Sridhar}, \&
  {Metzger}}]{Margalit:2020a}
{Margalit}, B., {Beniamini}, P., {Sridhar}, N., \& {Metzger}, B.~D. 2020, arXiv
  e-prints, arXiv:2005.05283.
\newblock \doarXiv{2005.05283}

\bibitem[{{Niino}(2018)}]{Niino:2018b}
{Niino}, Y. 2018, \apj, 858, 4 (N18), \dodoi{10.3847/1538-4357/aab9a9}

\bibitem[{{Niino} {et~al.}(2018){Niino}, {Tominaga}, {Totani}, {Morokuma},
  {Keane}, {Possenti}, {Sugai}, \& {Yamasaki}}]{Niino:2018a}
{Niino}, Y., {Tominaga}, N., {Totani}, T., {et~al.} 2018, \pasj, 70, L7,
  \dodoi{10.1093/pasj/psy102}

\bibitem[{{Palaniswamy} {et~al.}(2018){Palaniswamy}, {Li}, \&
  {Zhang}}]{Palaniswamy:2018a}
{Palaniswamy}, D., {Li}, Y., \& {Zhang}, B. 2018, \apjl, 854, L12,
  \dodoi{10.3847/2041-8213/aaaa63}

\bibitem[{{Petroff} {et~al.}(2015){Petroff}, {Bailes}, {Barr}, {Barsdell},
  {Bhat}, {Bian}, {Burke-Spolaor}, {Caleb}, {Champion}, {Chandra}, {Da Costa},
  {Delvaux}, {Flynn}, {Gehrels}, {Greiner}, {Jameson}, {Johnston}, {Kasliwal},
  {Keane}, {Keller}, {Kocz}, {Kramer}, {Leloudas}, {Malesani}, {Mulchaey},
  {Ng}, {Ofek}, {Perley}, {Possenti}, {Schmidt}, {Shen}, {Stappers}, {Tisserand
  }, {van Straten}, \& {Wolf}}]{Petroff:2015a}
{Petroff}, E., {Bailes}, M., {Barr}, E.~D., {et~al.} 2015, \mnras, 447, 246,
  \dodoi{10.1093/mnras/stu2419}

\bibitem[{{Petroff} {et~al.}(2016){Petroff}, {Barr}, {Jameson}, {Keane},
  {Bailes}, {Kramer}, {Morello}, {Tabbara}, \& {van Straten}}]{Petroff:2016a}
{Petroff}, E., {Barr}, E.~D., {Jameson}, A., {et~al.} 2016, \pasa, 33, e045,
  \dodoi{10.1017/pasa.2016.35}

\bibitem[{{Piro} \& {Burke-Spolaor}(2017)}]{Piro:2017a}
{Piro}, A.~L., \& {Burke-Spolaor}, S. 2017, \apjl, 841, L30,
  \dodoi{10.3847/2041-8213/aa740d}

\bibitem[{{Platts} {et~al.}(2019){Platts}, {Weltman}, {Walters}, {Tendulkar},
  {Gordin}, \& {Kandhai}}]{Platts:2019a}
{Platts}, E., {Weltman}, A., {Walters}, A., {et~al.} 2019, \physrep, 821, 1,
  \dodoi{10.1016/j.physrep.2019.06.003}

\bibitem[{{Popov} \& {Postnov}(2013)}]{Popov:2013a}
{Popov}, S.~B., \& {Postnov}, K.~A. 2013, ArXiv e-prints.
\newblock \doarXiv{1307.4924}

\bibitem[{{Prochaska} \& {Zheng}(2019)}]{Prochaska:2019a}
{Prochaska}, J.~X., \& {Zheng}, Y. 2019, \mnras, 485, 648,
  \dodoi{10.1093/mnras/stz261}

\bibitem[{{Prochaska} {et~al.}(2019){Prochaska}, {Macquart}, {McQuinn},
  {Simha}, {Shannon}, {Day}, {Marnoch}, {Ryder}, {Deller}, {Bannister},
  {Bhandari}, {Bordoloi}, {Bunton}, {Cho}, {Flynn}, {Mahony}, {Phillips},
  {Qiu}, \& {Tejos}}]{Prochaska:2019b}
{Prochaska}, J.~X., {Macquart}, J.-P., {McQuinn}, M., {et~al.} 2019, Science,
  366, 231, \dodoi{10.1126/science.aay0073}

\bibitem[{{Ravi} {et~al.}(2019){Ravi}, {Catha}, {D'Addario}, {Djorgovski},
  {Hallinan}, {Hobbs}, {Kocz}, {Kulkarni}, {Shi}, {Vedantham}, {Weinreb}, \&
  {Woody}}]{Ravi:2019a}
{Ravi}, V., {Catha}, M., {D'Addario}, L., {et~al.} 2019, \nat, 572, 352,
  \dodoi{10.1038/s41586-019-1389-7}

\bibitem[{{Shannon} {et~al.}(2018){Shannon}, {Macquart}, {Bannister}, {Ekers},
  {James}, {Os{\l}owski}, {Qiu}, {Sammons}, {Hotan}, {Voronkov}, {Beresford},
  {Brothers}, {Brown}, {Bunton}, {Chippendale}, {Haskins}, {Leach},
  {Marquarding}, {McConnell}, {Pilawa}, {Sadler}, {Troup}, {Tuthill},
  {Whiting}, {Allison}, {Anderson}, {Bell}, {Collier}, {G{\"u}rkan}, {Heald},
  \& {Riseley}}]{Shannon:2018a}
{Shannon}, R.~M., {Macquart}, J.~P., {Bannister}, K.~W., {et~al.} 2018, \nat,
  562, 386, \dodoi{10.1038/s41586-018-0588-y}

\bibitem[{{Staveley-Smith} {et~al.}(1996){Staveley-Smith}, {Wilson}, {Bird},
  {Disney}, {Ekers}, {Freeman}, {Haynes}, {Sinclair}, {Vaile}, {Webster}, \&
  {Wright}}]{Staveley-Smith:1996a}
{Staveley-Smith}, L., {Wilson}, W.~E., {Bird}, T.~S., {et~al.} 1996, \pasa, 13,
  243

\bibitem[{{Tendulkar} {et~al.}(2017){Tendulkar}, {Bassa}, {Cordes}, {Bower},
  {Law}, {Chatterjee}, {Adams}, {Bogdanov}, {Burke-Spolaor}, {Butler},
  {Demorest}, {Hessels}, {Kaspi}, {Lazio}, {Maddox}, {Marcote}, {McLaughlin},
  {Paragi}, {Ransom}, {Scholz}, {Seymour}, {Spitler}, {van Langevelde}, \&
  {Wharton}}]{Tendulkar:2017a}
{Tendulkar}, S.~P., {Bassa}, C.~G., {Cordes}, J.~M., {et~al.} 2017, \apjl, 834,
  L7, \dodoi{10.3847/2041-8213/834/2/L7}

\bibitem[{{The CHIME/FRB Collaboration}(2020){The CHIME/FRB
  Collaboration}, {:}, {Andersen}, {Band ura}, {Bhardwaj}, {Bij}, {Boyce},
  {Boyle}, {Brar}, {Cassanelli}, {Chawla}, {Chen}, {Cliche}, {Cook},
  {Cubranic}, {Curtin}, {Denman}, {Dobbs}, {Dong}, {Fandino}, {Fonseca},
  {Gaensler}, {Giri}, {Good}, {Halpern}, {Hill}, {Hinshaw}, {H{\"o}fer},
  {Josephy}, {Kania}, {Kaspi}, {Landecker}, {Leung}, {Li}, {Lin}, {Masui},
  {Mckinven}, {Mena-Parra}, {Merryfield}, {Meyers}, {Michilli}, {Milutinovic},
  {Mirhosseini}, {M{\"u}nchmeyer}, {Naidu}, {Newburgh}, {Ng}, {Patel}, {Pen},
  {Pinsonneault-Marotte}, {Pleunis}, {Quine}, {Rafiei-Ravandi}, {Rahman},
  {Ransom}, {Renard}, {Sanghavi}, {Scholz}, {Shaw}, {Shin}, {Siegel}, {Singh},
  {Smegal}, {Smith}, {Stairs}, {Tan}, {Tendulkar}, {Tretyakov}, {Vanderlinde},
  {Wang}, {Wulf}, \& {Zwaniga}}]{CHIME:2020a}
{The CHIME/FRB Collaboration}, {:}, {Andersen}, B.~C., {et~al.} 2020, arXiv
  e-prints, arXiv:2005.10324.
\newblock \doarXiv{2005.10324}

\bibitem[{{Thornton} {et~al.}(2013){Thornton}, {Stappers}, {Bailes},
  {Barsdell}, {Bates}, {Bhat}, {Burgay}, {Burke-Spolaor}, {Champion}, {Coster},
  {D'Amico}, {Jameson}, {Johnston}, {Keith}, {Kramer}, {Levin}, {Milia}, {Ng},
  {Possenti}, \& {van Straten}}]{Thornton:2013a}
{Thornton}, D., {Stappers}, B., {Bailes}, M., {et~al.} 2013, Science, 341, 53,
  \dodoi{10.1126/science.1236789}

\bibitem[{{Tominaga} {et~al.}(2018){Tominaga}, {Niino}, {Totani}, {Yasuda},
  {Furusawa}, {Tanaka}, {Bhand ari}, {Dodson}, {Keane}, {Morokuma}, {Petroff},
  \& {Possenti}}]{Tominaga:2018a}
{Tominaga}, N., {Niino}, Y., {Totani}, T., {et~al.} 2018, \pasj, 70, 103,
  \dodoi{10.1093/pasj/psy101}

\bibitem[{{Totani}(2013)}]{Totani:2013a}
{Totani}, T. 2013, \pasj, 65, L12, \dodoi{10.1093/pasj/65.5.L12}

\bibitem[{{Walker} {et~al.}(2018){Walker}, {Ma}, \& {Breton}}]{Walker:2018a}
{Walker}, C.~R.~H., {Ma}, Y.~Z., \& {Breton}, R.~P. 2018, arXiv e-prints,
  arXiv:1804.01548.
\newblock \doarXiv{1804.01548}

\bibitem[{{Xu} \& {Han}(2015)}]{Xu:2015a}
{Xu}, J., \& {Han}, J.~L. 2015, Research in Astronomy and Astrophysics, 15,
  1629, \dodoi{10.1088/1674-4527/15/10/002}

\bibitem[{{Yamasaki} \& {Totani}(2020)}]{Yamasaki:2020a}
{Yamasaki}, S., \& {Totani}, T. 2020, \apj, 888, 105,
  \dodoi{10.3847/1538-4357/ab58c4}

\bibitem[{{Yang} {et~al.}(2017){Yang}, {Luo}, {Li}, \& {Zhang}}]{Yang:2017a}
{Yang}, Y.-P., {Luo}, R., {Li}, Z., \& {Zhang}, B. 2017, \apjl, 839, L25,
  \dodoi{10.3847/2041-8213/aa6c2e}

\bibitem[{{Zhang}(2017)}]{Zhang:2017a}
{Zhang}, B. 2017, \apjl, 836, L32, \dodoi{10.3847/2041-8213/aa5ded}

\end{thebibliography}

\end{document}